\documentclass[11pt]{article}
\usepackage{amsmath,amsfonts}
\usepackage[numbers,sort&compress]{natbib}
\setcitestyle{aysep={}}
\usepackage[left=2.54cm,top=2.54cm,right=2.54cm,bottom=2.54cm,bindingoffset=0.0cm]{geometry}
\usepackage{setspace}
\usepackage{parskip} 
\usepackage[left]{lineno}
\usepackage{hyperref}
\usepackage{graphicx}
\usepackage{amssymb}
\usepackage{float}
\usepackage[labelfont=bf]{caption} 
\usepackage{authblk} 
\usepackage{xcolor}
\usepackage[utf8]{inputenc}
\usepackage{amsmath}
\usepackage{booktabs}
\usepackage{graphicx}
\usepackage{overpic}
\usepackage{rotating}
\usepackage{pdflscape}
\usepackage{longtable}
\usepackage{rotating}

\usepackage{fancyhdr}
\newcommand{\beginsupplement}{%
        \setcounter{table}{0}
        \renewcommand{\thetable}{S\arabic{table}}%
        \setcounter{figure}{0}
        \renewcommand{\thefigure}{S\arabic{figure}}%
     }

\begin{document}

\title{Aggression heuristics underlie animal dominance hierarchies and provide evidence of group-level social information}

\fancyhead[L]{Hobson et al.}
\fancyhead[R]{\textit{Aggression heuristics and social information}}



\author[a,b,*]{Elizabeth A. Hobson}
\author[c,d,e]{Dan M{\o}nster}
\author[b,f]{Simon DeDeo}

\affil[a]{Department of Biological Sciences, University of Cincinnati, Cincinnati, OH USA}
\affil[b]{Santa Fe Institute, Santa Fe, NM USA}
\affil[c]{Interacting Minds Centre, Aarhus University, Aarhus C, Denmark}
\affil[d]{School of Business and Social Sciences, Aarhus University, Aarhus V, Denmark}
\affil[e]{Cognition and Behavior Lab, Aarhus University, Aarhus V, Denmark}
\affil[f]{Social and Decision Sciences, Dietrich College, Carnegie Mellon University, Pittsburgh, PA USA}
\affil[*]{Corresponding author: elizabeth.hobson@uc.edu}  
\date{}
\maketitle

\textbf{Abstract}

Members of a social species need to make appropriate decisions about who, how, and when to interact with others in their group. However, it has been difficult for researchers to detect the inputs to these decisions and, in particular, how much information individuals actually have about their social context. We present a new method that can serve as a social assay to quantify how patterns of aggression depend upon information about the ranks of individuals within social dominance hierarchies. Applied to existing data on aggression in 172 social groups across 85 species in 23 orders, it reveals  three main patterns of rank-dependent social dominance: the \textit{downward heuristic} (aggress uniformly against lower-ranked opponents), \textit{close competitors} (aggress against opponents ranked slightly below self), and \textit{bullying} (aggress against opponents ranked much lower than self). The majority of the groups (133 groups, 77\%) follow a downward heuristic, but a significant minority (38 groups, 22\%) show more complex social dominance patterns (close competitors or bullying) consistent with higher levels of social information use. These patterns are not phylogenetically constrained and different groups within the same species can use different patterns, suggesting that heuristics use may depend on context and the structuring of aggression by social information should not be considered a fixed characteristic of a species. Our approach provides new opportunities to study the use of social information within and across species and the evolution of social complexity and cognition.  






\singlespacing

\section*{Introduction}

Social individuals can gain from social interactions, but close associations with potential competitors introduce the risk of costly aggression. Decisions about who and how to interact with others are often based on an individual's assessment of its own abilities (\textit{e.g.}~\citep{Hobson2020DifferencesHierarchies, chase1994aggressive, Landau1951}) or on the strength of relationships between pairs of individuals. These decisions can also be made on the basis of a larger social context---most notably, on the basis of rank in a dominance hierarchy. Evidence across multiple social situations shows that decisions about interactions can be affected by rank across both humans and animals: friendships among school children~\citep{Ball2013FriendshipStatus}, messaging patterns in online social dating apps~\citep{Bruch2018AspirationalMarkets}, and grooming in female primates (\emph{e.g.}, ~\citep{Henzi1999ThePrimates, Seyfarth1977AMonkeys}) can all be affected by rank.

Almost 100 years of research on animal conflict has shown that rank matters. Dominance hierarchies structure group interactions in a vast array of animals, from primates and hyenas to fish and wasps (\textit{e.g.},~\citep{SchjelderupEbbe1922,Vehrencamp1983,Shizuka2012,Strauss2019SocialSocieties,Grosenick2007FishAlone,tibbetts2004}), including humans (\textit{e.g.},~\citep{mascaro2014,mascaro2012}). Previous research has shown that aggression networks underlying dominance hierarchies in species across the phylogenetic tree are built from remarkably similar basic structures~\citep{Shizuka2015}. Other studies have documented the large effects rank can have on an individual's stress, health, and fitness (\textit{e.g.},~\citep{sapolsky2004social,sapolsky2005influence,Creel2001SocialHormones}). In the last 30--70 years, studies in both empirical and theoretical contexts have provided insight into the major factors affecting the formation of hierarchies (\textit{e.g.}, ~\citep{Chase1982,Landau1951OnCharacteristics,Dugatkin1997}). 

Although we now understand that dominance hierarchies are widespread, that rank is often important, and the basics of how hierarchies form in many species, a critical open question is what animals within these hierarchies ``know'' about their own rank and the ranks of others. Social information is increasingly recognized as a critical component for understanding the structure of animal societies~\citep{Hobson2019RethinkingConcepts,Seyfarth2010TheCommunication}. Individuals can gather social information by attending to the signals and behaviors of their group members ~\citep{Page2019TheBats,Danchin2004PublicEvolution,Hobson2020DifferencesHierarchies}. If individuals can perceive something about their own rank or the ranks of others in their group, they could use that information to better maximize their potential gains from aggression and minimize potential losses or injury. In the context of conflict in hierarchically-ordered groups, various kinds of social information can be gleaned from the outcomes of aggressive interactions such as social information about an individual's own ability to win fights against opponents, the relationships it has with others, relationships among others in the group, an individual's own rank and the rank of others, or the group's overall dominance structure~\citep{Hobson2020DifferencesHierarchies}. The more information that individuals can access, process, and use in their decision-making, the more patterns of micro-level aggressive actions become fundamentally entwined with macro-level structural information about rank in social groups~\citep{Hobson2019RethinkingConcepts}. 

Most previous studies to detect social information about rank in animals have required extensive experimental manipulation, involving reversing the apparent outcome of observed fights and testing whether uninvolved individuals are more attentive to fights which violate the order of rank in the hierarchy than fights with more expected outcomes. These experiments have been instrumental in demonstrating the extent of rank information contained in some animal groups (\textit{e.g.}~\citep{Bergman2003,Cheney1995}). However, it has been difficult to assess rank information across many species because these experiments are time intensive and require social systems in which fight outcomes can be easily artificially manipulated, which limits our abilities to conduct comparative analyses of information use across animals.

We take a new approach to address the question of what animals may ``know'' about rank by quantifying the presence, amount, and type of \textit{social information} contained in animal dominance hierarchies. We define social information as any information about an individual's interactions, relationships, or status, held by that individual about itself or others in its group. To quantify social information, we developed computational methods to detect signatures of the kinds of social dominance patterns that characterize conflict in animal hierarchies. Our approach has three major benefits: (1) it provides computational rather than experimental methods that allow for detection of the presence and use of information, (2) the methods can be used with  existing data, providing new opportunities for comparisons across across a wide range of species, and (3) our focus on the structural properties and how information is contained and used in animal systems is agnostic to whether emergent patterns are based on complex cognition and strategic decision-making, or are the result of much simpler mechanistic rules. 

Our approach focuses on inferring the kinds of information about rank that are contained within patterns of social interactions. Our methods allow us to connect each individual's micro-level decisions about aggression with macro-level social properties like the structure of group dominance hierarchies. If the same decision-making process is used across individuals in a group, a group's aggression patterns can be characterized. We refer to \textit{rank-dependent aggression} as conflict in animals groups that is contingent on the relative rank differences between the individuals. Rank-dependent aggression forms the basis for the emergence of simple rules or heuristics about aggression, which we refer to as \textit{social dominance patterns}. Different social dominance patterns may emerge depending on the detail of rank information individuals have. 

We apply these methods to a large empirical dataset on aggression and dominance in 172 independent social groups across 85 species in 23 orders ~\cite{Shizuka2015,hobsondedeo2015}. 

To detect rank-dependent social dominance patterns, we developed a four-step process. First, we developed \textit{focus} and \textit{position} as new summary measures to quantify the extent to which group conflict is affected by rank. Each individual in the group is assessed to determine how it aggresses against opponents based on relative rank difference (how many steps in rank above or below the aggressor its opponents are in rank). \textit{Focus} quantifies the extent to which aggression is concentrated on a subset of opponents and measures the fraction of aggression that \textit{was} directed between individuals separated a certain number of steps in relative rank compared to aggression that \textit{could have been} directed the same number of steps away. If rank information is present and is used to concentrate aggression, then knowing where in relative-rank difference terms the peak of aggression is focused can tell us about the kind of social dominance pattern the group is using. To differentiate between different ways that rank may inform decision-making, we measure a second quantity, \textit{position}, which reflects \textit{where} in relative rank difference aggressors concentrate their aggression. 

We then defined three main social dominance patterns: (1) the \textit{downward heuristic}, where individuals aggress against lower-ranked individuals regardless of their particular rank value relative to the aggressor; (2) \textit{close competitors}, where individuals aggress preferentially towards those just below themselves in rank; and (3) \textit{bullying}, where individuals aggress preferentially towards those ranked far below themselves in rank. Next, we categorized which social dominance pattern animals in each group followed. We assigned social dominance type by comparing focus and position values from the observed groups to those produced by an ensemble of permutation-based \textit{reference models}~\citep{Hobson2020refmodels, Gauvin2018} simulating conflict via specified rules. These reference models allow us to simulate what aggression \textit{should} look like if individuals in the group only follow the specified interaction rules rather than incorporating any additional information about the ranks of their opponents.

Finally, we compared the reference datasets to the empirical datasets to evaluate whether observed aggression patterns could plausibly have been generated by animals following the simplest social dominance pattern (the downward heuristic) or if more detailed information is needed to describe observed aggression patterns. Importantly, our methods are agnostic to the ways in which social information is encoded in these social systems. Information could be stored cognitively, but may also be encoded in other less cognitively-demanding ways, such as through observable signals. 

Combined, our new quantitative methods, our reference model comparisons, and our detection of social rules governing social dominance patterns within hierarchies provide new insight into how animals structure their social relationships and how they make biologically-relevant social decisions.

\section*{Results and Discussion}

\subsection*{Rank-dependent social dominance}

\begin{figure} [ht]
\centering
\caption{\textit{Focus} and \textit{position} values for observed social groups, colored by social dominance pattern type (see Fig.~\ref{strategy.exs} for categorization). Focus is a measure of how concentrated aggression is, given the relative rank differences from all individuals to their potential opponents (as more aggression is restricted to a subset of opponents focus values become higher). Position values measure where in relative rank difference this aggression is concentrated: when individuals focus their aggression on opponents ranked just below themselves in the hierarchy, position values are near 0; when aggression is focused on opponents ranked far below, position is near 1.  }
\label{FxP}
\includegraphics[width=.65\textwidth]{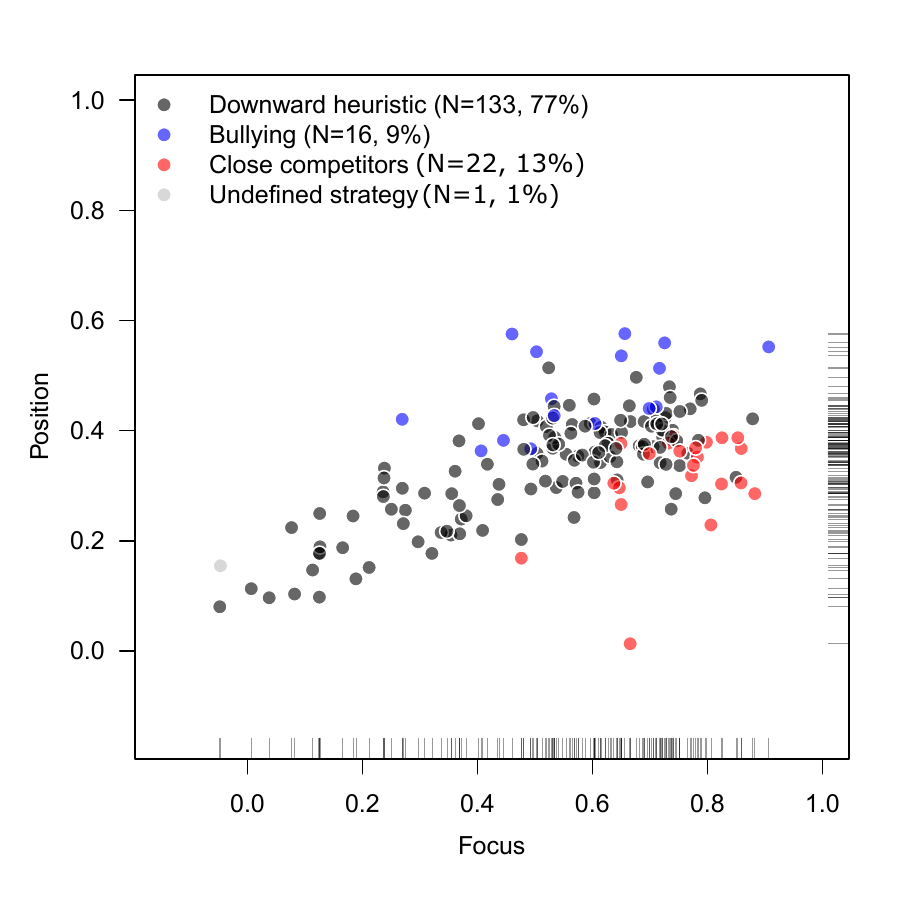} 
\end{figure}

\begin{figure} 
\centering
\caption{Each social group was categorized by which rank-dependent social dominance pattern they followed. Shown here are three examples of pattern assignment, to (a) \textit{downward heuristic} (mule deer,~\citep{koutnik1981}), (b) \textit{close competitors} (monk parakeet,~\citep{hobsondedeo2015}), and (c) \textit{bullying} (vervet monkey,~\citep{isbell1998}). Diamond points show observed \textit{focus} and \textit{position} values for each group. Grey circular points indicate focus and position values ($\pm$ 95\% CI) for reference model datasets generated using a downward heuristic with different proportions of randomly-directed aggression (see inset). This ensemble of reference models shows how expected focus and position values change as the proportion of randomly-directed aggression events increase, from no randomly-directed aggression (100\% adherence to the downward heuristic with all aggression directed towards lower-ranked opponents, right side) to fully randomly-directed aggression (left side, where aggression is purely driven by individual aggressiveness, with no rank information). Social dominance patterns for each group were assigned by comparing focus and position values in each empirical group to the reference model ensemble for that group. If the observed value fell within the downward heuristic polygon, empirically-observed focus and position values could have been produced by a downward heuristic; when values fell outside this polygon, another pattern is needed to explain the observed empirical patterns. When position values were lower than expected, and aggression towards opponents ranked close below in the hierarchy was more common, we categorized the aggression pattern as close competitors and when position values were higher than expected, and aggression toward opponents ranked far below in the hierarchy was more common, we categorized the pattern as bullying.}
\label{strategy.exs}
\includegraphics[width=1.0\textwidth]{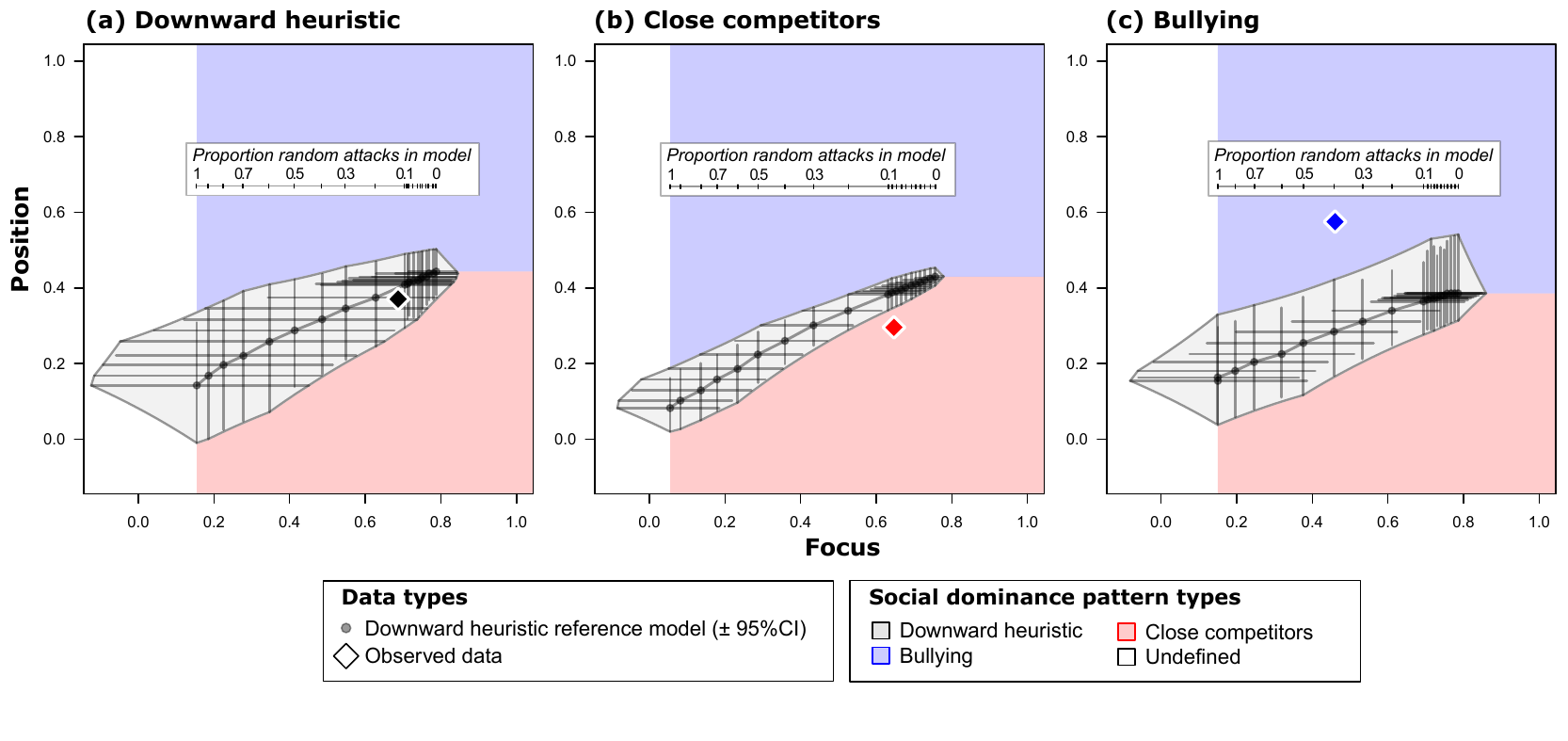} 
\end{figure}

Our measures of \textit{focus} show that the majority of animal social groups in our dataset had evidence of structured aggression (see Supplementary Information~\ref{SI-strucagg} and~\ref{SI-data}). While focus values show how strong the hierarchical organization is, our measures of \textit{position} allow us to diagnose the type of social dominance individuals used within the hierarchical structure. Based on our summary measures of \textit{focus} and \textit{position} for each group (Fig.~\ref{FxP}), aggression patterns in nearly all groups ($99\%$ of groups, $N=171$) could be categorized without ambiguity to one of three main aggression patterns: the \textit{downward heuristic}, \textit{close competitors}, or \textit{bullying} (Fig.~\ref{strategy.exs}). 

Each social dominance pattern emerges as individuals preferentially engage with a certain subset of opponents and may be based on the level of detail individuals have about the relative rank difference between themselves and potential opponents. The downward heuristic is the simplest of the three main social dominance patterns, and emerges when individuals aggress indiscriminately towards lower-ranked opponents. The focus and position values in $77\%$ of empirical datasets ($N=133$) could have been produced by animals following a simple downward heuristic. However, $22\%$ (38 groups) used social dominance patterns where additional rank information is needed in order to produce the observed patterns. We classify both \textit{close competitors} and \textit{bullying} as more complex social dominance patterns because they are based on more detailed rank information than the downward heuristic, as aggressors need to differentiate between lower-ranked potential opponents by whether they are ranked just below or far below themselves in the hierarchy. A \textit{close competitors} aggression pattern (preferentially aggress against opponents ranked slightly below themselves) was used by $13\%$ of groups and a \textit{bullying} pattern (preferentially aggress against opponents ranked far below themselves) was used by $9\%$ of groups. Only $1$ group had an \textit{undefined} social dominance pattern. 

\subsection*{Evaluating other potential generative processes}
Within our empirical datasets on aggression, we found no evidence that a group's social dominance pattern use could be consistently explained by the number of individuals in the social system or whether the group was observed in natural conditions or captivity (see~\ref{SI-factors}, Fig.~\ref{ridgeplot_grptype} and Fig.~\ref{ridgeplot_grpsize}). 

We use the presence of either a close competitor or bullying social dominance pattern as an indication of higher levels of social information. These patterns may emerge if individuals have access to that social information and use it to structure their fights with particular opponents beyond simply reacting to their own experiences and treating opponents as interchangeable or anonymous. However, individuals (especially across very different species, with different cognitive systems) may not have access to this more detailed social information~\citep{Hobson2020DifferencesHierarchies}. 

To investigate the role of information in the emergence of more complex social dominance patterns, we constructed another model of aggression to determine how often more complex social dominance patterns might emerge when information is more limited (see~\ref{SI-WLeffects}). We used a generative-process reference model~\cite{Hobson2020refmodels} to simulate social groups with 10 individuals to examine how individual-level information about wins and losses result in the emergence of group-level social dominance patterns in the absence of the ability to collect information about the ranks of others in the group. Each individual in our model only has access to its own win/loss record, and can only adjust its behavior based on outcomes of events (individuals do not have any information about which other individuals they interacted with or which individuals they have won or lost against). We modeled nine variants: a winner-effect only model, a loser-effect only model, and a mixed winner and loser effect model; each of these models was further investigated using a transient effect and a permanent effect (Table~\ref{tab:effect_combos}), using winner and loser effect strengths from the literature~\citep{Rutte2006WhatLosing} along with both a more extreme and a more moderate value for comparison. 

Across all model variants, the majority of simulated groups showed aggression consistent with the downward heuristic social dominance pattern, demonstrating that basic hierarchical group structures can be produced when social information is limited. However, simulated group aggression rarely resulted in a bullying or close competitors pattern when information was limited (Table~\ref{tab:strat_freq}). This pattern is even more apparent when we focused on ``realistic'' winner and loser effect values~\citep{Rutte2006WhatLosing} and excluded simulated groups that differed strongly in structure from our empirical datasets (i.e., focus and/or position were less than 0, Fig.~\ref{fig:Rutte_FP}): bullying patterns were then only observed in 0\% (transient effects) and 2.3\% (persistent effects) and close competitors were only observed in 1.8\% (transient effects) and 1.14\% (persistent effects) (see Table~\ref{tab:Rutte}). These results show that although it is possible to produce a close competitor or bullying social dominance pattern with individual-level information only, it is rare for these more complex patterns to emerge in the absence of additional social information. This is additional evidence for treating close competitor and bullying patterns as more information-rich patterns than the downward heuristic, and likely information beyond individual experience is required to reliably produce close competitor or bullying patterns.

\subsection*{Phylogenetic signal and the evolution of rank-dependent aggression}

\begin{figure}
\centering
\caption{Social dominance pattern types are not phylogenetically restricted to particular orders. Studies of aggression in animals included in our dataset are unevenly distributed across orders (a) as well as whether multiple groups of a particular species have been sampled (b, see also Fig.~\ref{fig:mxspp}). When these totals are broken down by aggression pattern type (c), we see that many orders have groups with more than one aggression type (number of groups listed in table, percent of groups by aggression pattern for each order indicated by color code, where red indicates 100\% of sampled groups showed a particular type). In most cases, groups within many orders did not have consistently simple (downward heuristic) or consistently complex (bullying or close competitors) aggression patterns (d). Note: Ovalentaria is a group of fish families categorized as \textit{incertae sedis} (`of uncertain placement').}
\label{StratsxOrder}
\includegraphics[width=0.9\textwidth]{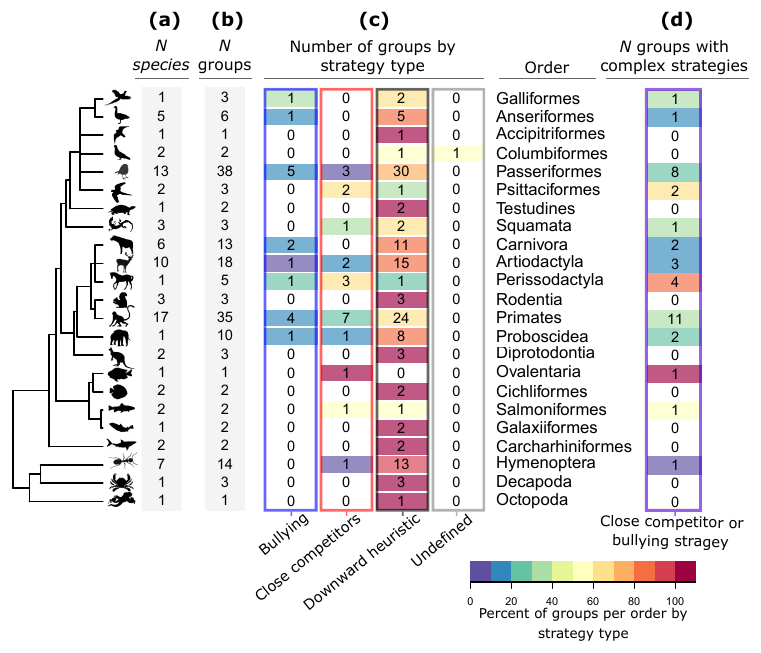} 
\end{figure}

\begin{figure} 
\centering
\caption{Little to no evidence of evolutionary relatedness on social dominance patterns is present at the order level. In almost every order, the observed number of groups with each pattern (solid vertical lines) overlaps with the number of groups with each pattern when patterns are randomly allocated (shaded areas, density estimates) for each of the three main social dominance patterns (a) downward heuristic, (b) close competitors, and (c) bullying. Note: Ovalentaria is a group of fish families categorized as \textit{incertae sedis} (`of uncertain placement'). } 
\label{ridgeplot_order}
\includegraphics[width=0.99\textwidth]{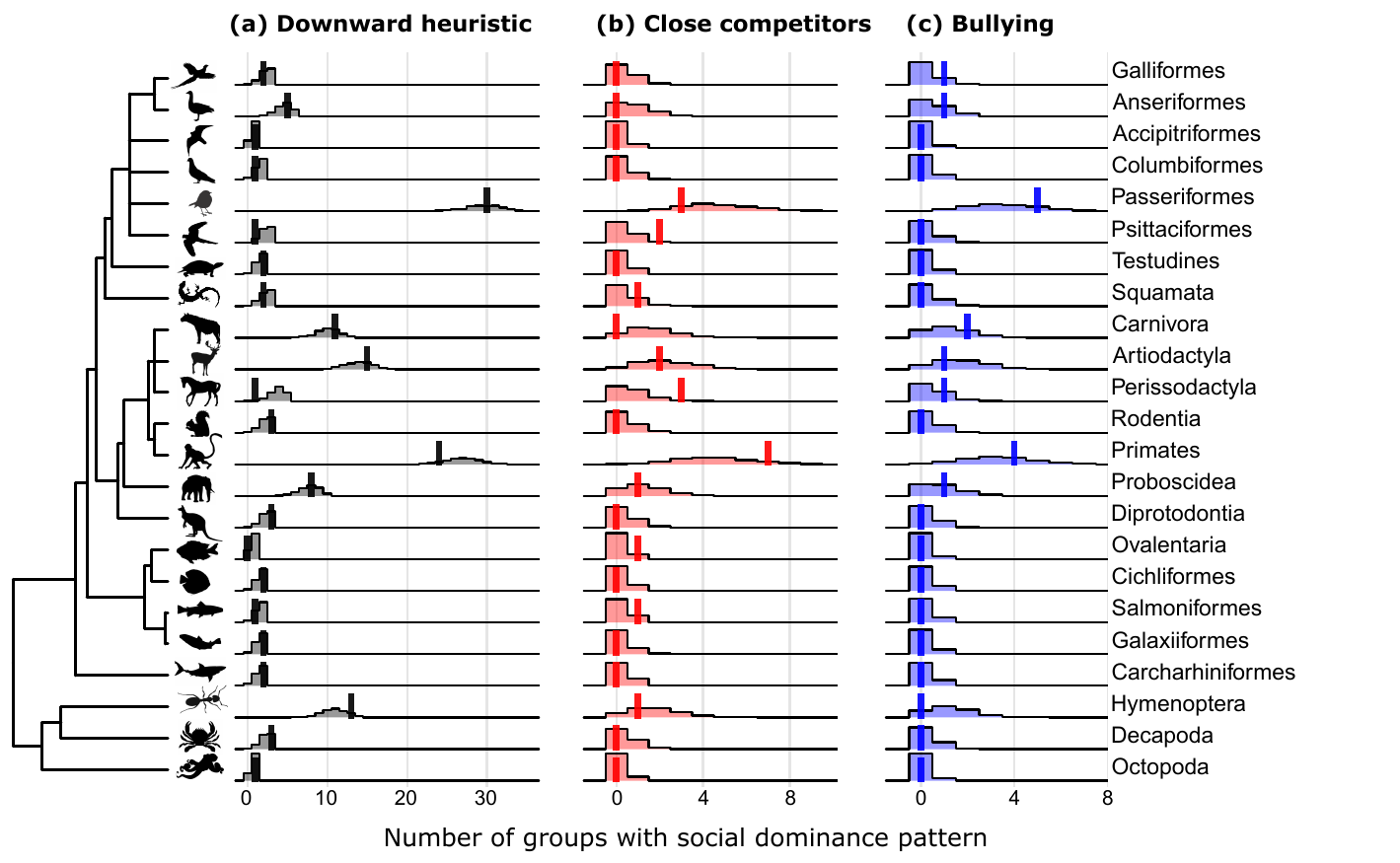}
\end{figure}

All three well-defined aggression patterns occurred in orders across the range of animal groups in our dataset of 172 social groups across 85 species and 23 orders (Fig.~\ref{StratsxOrder}). We found no consistent evidence for phylogenetic signal in the evolution of any of the three social dominance patterns: the frequency at which each pattern occurred within each of the orders was consistent with the distribution expected from random allocation of patterns in almost every case (Fig.~\ref{ridgeplot_order}). With $\alpha=0.025$ for two-tailed tests, Perissodactyla was the only order where the observed number of groups differed significantly from the randomized frequencies: the occurrence of downward heuristic social dominance patterns was lower than expected if patterns are randomly distributed ($p=0.012$). Both Perissodactyla and Psittaciformes showed some evidence of unusually higher frequencies of observed close competitors patterns ($p=0.026$ and $p=0.037$, respectively). Due to the many comparisons shown here, these results should be interpreted with caution, but are indications that future studies of species in these orders is warranted. 

Although certain kinds of conflict can be associated with phylogenetic relatedness, such as the occurrence of lethal violence in mammals~\citep{gomez2016phylogenetic}) or the steepness of dominance hierarchies within a clade of primates~\citep{balasubramaniam2012hierarchical}, other studies have found more consistency in aggression and dominance across species. For example, studies of the structure and frequency of network motifs within aggression networks have found striking similarities across species at the micro social scale~\citep{Shizuka2015}. Our work builds on these previous findings, although we take a complementary approach by addressing hierarchical structures from a macro-structural perspective. Rather than focusing at the building blocks of hierarchies, we looked at the social dominance patterns that may underlie aggression decisions. However, even coming at this question from the opposite scale, we find similar patterns, where macro level structures cannot be explained by phylogenetic relatedness. It is important to note that these historical datasets are taxonomically biased towards overrepresentation of certain clades (e.g., birds and primates) and an underrepresentation of studies in many others (see Fig.~\ref{StratsxOrder}a). Future work on a broader range of species will provide more balanced insight into evolutionary patterns. 

Our methods allow us to detect social information within groups that could form the basis for simple heuristics to guide aggression, but cannot differentiate between the availability or \textit{presence} of information and the intentional \textit{use} of that information. Although animals may vary widely in their underlying perception, memory, inference, and recognition skills, our results show that the social information contained in groups can be used to structure aggression. This raises the possibility the same social dominance patterns may emerge from very different cognitive mechanisms, decision-making heuristics, or social information processing abilities. Manipulative experiments are needed in order to differentiate the types of processes that generate and store information in high-information social groups. These experiments will provide valuable insight into whether the emergence of these patterns is indicative of more complex social strategies, based on cognitive processing, strategic decision-making, and flexible social competence, or whether these patterns can be explained by simpler rules. Simple rules, or \textit{heuristics}~\citep{tversky1974judgment,Slovic1977}, are a major factor that structures human social behavior and decision making and characterizing these heuristics and the advantages and disadvantages of their use has allowed economists and psychologists to explain previously mystifying features of human behavior (\emph{e.g.},~\citep{hertwig2013simple, nagatsu2018making}). A better understanding of the kinds of heuristics animals may use to make decisions and the ways animals may respond to changing social conditions by altering their heuristics has the potential to provide new and valuable insight into animal social complexity.

\subsection*{Intraspecific variation in aggression pattern use}

\begin{figure} 
\centering
\caption{Occurrence of social dominance patterns by species with multiple empirically-sampled groups. In almost every species with multiple groups, the observed number of groups with each pattern (solid vertical lines) overlaps with the number of groups with each pattern when patterns are randomly allocated (shaded areas, density estimates) for each of the three main social dominance patterns (a) downward heuristic, (b) close competitors, and (c) bullying. Asterisks indicate species with unusually low or high numbers of observed social dominance patterns compared to the randomized patterns ($\alpha<0.025$ for two-tailed test; species with $\alpha<0.05$ indicated with annotated p-values). Data are sorted by number of groups then by pattern type.}
\label{fig:mxspp}
\includegraphics[width=1.0\textwidth]{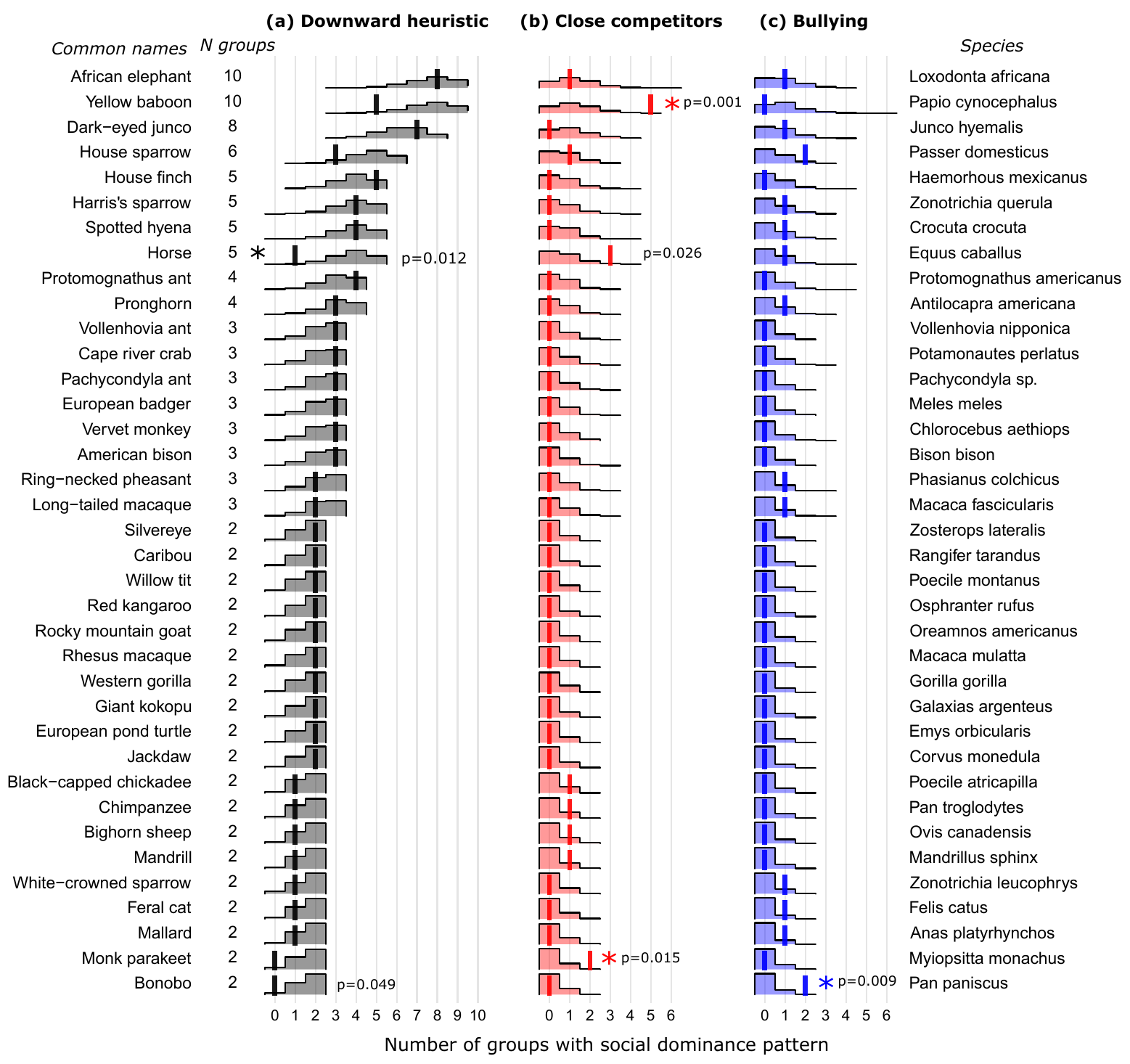} 
\end{figure}

While the social dominance patterns used by different groups were sometimes consistent within a species, we found multiple cases where different groups followed different rank-dependent patterns. For the $37$ species for which two or more groups were consistent with one of the three aggression patterns (downward heuristic, close competitors, or bullying), $46\%$ of species had groups that followed more than one aggression pattern (Fig.~\ref{fig:mxspp}). For example, yellow baboons were evenly split between 5 groups which used a basic downward heuristic and 5 groups that used the more complex close competitors pattern. Three species, African elephants, house sparrows, and horses, had groups that followed each of the three social dominance patterns. Of these species with multiple observed groups, horses and bonobos showed some evidence of a lower than expected frequency of downward heuristic patterns ($p=0.012$ and $p=0.049$ respectively, Fig.~\ref{fig:mxspp}). Yellow baboons, horses, and monk parakeets showed some evidence for higher than expected frequencies of close competitor dominance patterns ($p=0.001$, $p=0.026$, and $p=0.015$ respectively, Fig.~\ref{fig:mxspp}) and bonobos showed evidence of higher than expected frequencies of bullying social dominance ($p=0.009$, Fig.~\ref{fig:mxspp}). Care must be taken in interpreting these results due to multiple comparisons, but provide further indications of species which may be particularly interesting for future more detailed work.

This variability we find in which social dominance pattern occurs within species shows that these patterns should be thought of as facts about particular groups rather than rigid species-level characteristics. Factors such as resource availability and distribution, environmentally-mediated constraints, and direct environmental influences on physiology can all result in changes to individual aggression and group dominance structure (reviewed in~\citep{wong2012abiotic}). These changes may shift which aggression pattern is optimal under new social, environmental, or ecological conditions. Temporal shifts in the behaviors underlying dominance interactions have been documented in human groups where dominance patterns and the behaviors used to mediate dominance interactions change with age~\cite{Hawley1999}. Dominance patterns can even change over time within the same social group, as we previously documented in aggression in parakeets~\citep{hobsondedeo2015}. Our results support these earlier conclusions that sociality can vary within a single species. 

Combined, these results suggest that experimental work on the emergence and dynamics of dominance hierarchies, social information, and social dominance patterns is needed to fully understand the conditions under which an information-based aggression pattern, like rank-dependent aggression, would emerge and be used in social groups. In particular, more studies are needed to determine the range of social dominance patterns that a particular species is able to use, whether there are similarities in the social or environmental conditions under which a more information-rich pattern generally emerges, and how flexible and on what time scale pattern use may vary within a particular social group.

\section*{Conclusions}

A fundamental question in animal behavior is how much animals know about their social worlds and the extent to which they use this information in their decision-making processes~\citep{Hobson2019RethinkingConcepts,Hobson2020DifferencesHierarchies,Seyfarth2015SocialCognition,Seyfarth2010TheCommunication}. Many approaches to social complexity seek to understand how much animals know about their social worlds, and recent work has advocated explicitly quantifying social information when attempting to assess social complexity~\citep{Bergman2015,Hobson2019RethinkingConcepts}. There is growing evidence that social information is actively sought by individuals across a wide range of species~\citep{Barve2020TrackingWars,hobsondedeo2015,Hotta2015TheFish,Tibbetts2020WaspsRivals}. However, while we can quantify many aspects of social structure, without additional experimental manipulation (\textit{e.g.}~\citep{Bergman2003,Cheney1995}) it has not previously been possible to determine the extent of information that individuals in groups may have of their social worlds. In broader comparisons, it has also been difficult to find a way to quantify social information in a manner that is both feasible and general enough to be used in a wide range of species, as social interactions may differ in their salience and biological meaningfulness across species. 

The computational methods presented here provide a new way to assay interactions like aggression to determine the kinds of information encoded in social systems. We can now use these approaches to infer how much information animals have about their social worlds, based on their decisions about how to interact with each other. Using these new tools, researchers can now categorize groups into a taxonomy of social dominance patterns, where the structuring of aggression is based on different types of social information. 

The broad applicability of our quantitative tools provides new opportunities to quantify the evolution of social structure across divergent taxa and groups with many different types of social organization. The tractability and wide applicability of our approach enables comparative analyses that can provide a better understanding of the evolutionary patterns underlying the distribution of social processing skills and complex sociality across taxa. Combined with recent results from empirical work and an understanding of the cognitive abilities of species, our approach provides new opportunities to investigate the extent of rank-based information encoded in societies across species, compare the evolution of the use of social information, and better understand the effect of social information on individual behavior in within-group conflict.

The evidence we found for the role of social information in establishing social dominance patterns suggests that the question of what animals know about their social worlds should be thought of in two parts: first, ``how much do they know?''; second, ``how do they know it?''. Our results here deal with the extent of rank information animal groups have, but cannot determine the mechanisms through which rank becomes ``known'' by individuals. A better understanding of the cognitive abilities of the species, including memory, recognition, and perceptive abilities, is needed to fully understand how information is encoded and the kinds of cognition that underlie the entire process. For species that have more detailed information about rank and use a close competitors or bullying aggression pattern, priorities for future research will be to differentiate between cases where individuals can follow a more information-rich social dominance pattern via a simple underlying rule that allows easy detection of relative rank differences compared to cases where the ability to use rank information is based instead on more cognitively demanding methods that require the recognition of particular individuals and memories of past outcomes. Manipulative experiments are needed in order to differentiate the types of processes that generate and store information in high-information social groups. These kinds of experiments are critical in distinguishing between social groups where information is contained in more or less cognitively-demanding ways, and will allow us to begin to identify those species that could have more or less complex social assessment and memory abilities than commonly assumed.

\section*{Methods}
%

\subsection*{Empirical data sources}
We used a large openly accessible empirical dataset of aggression and dominance hierarchies (\cite{Shizuka2015}, \url{https://doi.org/10.5061/dryad.f76f2}). We excluded two of these datasets due to apparent errors in the presentation of data in the original papers (Table~4, Nest~39 in~\cite{Blatrix2004} and Table~3 in~\cite{Fairbanks1994}). We supplemented this dataset with data from aggression and rank in two groups of monk parakeets (\cite{hobsondedeo2015}, \url{https://doi.org/10.5061/dryad.p56q7}, data from study quarters 2-4 for groups 1 and 2). 

These datasets contain the number of times each individual ``won'' against each other individual. Depending on how the original study reported their data, ``wins'' could be the outcome of aggressive contests, show the directionality of aggressive events, or indicate a submission display towards a dominant individual (they do not have information on which individual started a fight, only the outcome of the interaction). We use the general term ``aggress'' to describe the actions individuals take in these datasets, and focus here on the perspective of the winners as initiators of aggression, although all of our analyses apply equally well to cases where the initiator of the fight chose to start a fight that it ultimately lost.

\subsection*{Rank and distribution of aggression}
For each group, we find individual ranks using a modified version of eigenvector centrality. In particular, we compute the probability that each individual aggresses with each other individual, and then add a small regularization term, $\epsilon$ (see SI Appendix~\ref{SI-epsilon} for a Bayesian calculation of  the optimal value of this term); the eigenvector centrality of the resulting matrix allows us to extract the relative ranks of individuals that are implicit in the patterns of aggression~\cite{brush2013family}.

Plotting the overall distribution of observed aggression in each group by relative rank differences enables us to determine whether the distribution of aggression is structured by rank differences among individuals, whether individuals in the group focus their aggression on a subset of individuals based on relative rank differences, and where in relative rank distance space aggression is focused. We quantify these characteristics by measuring \textit{focus} and \textit{position} (defined below). In the measurement of both quantities, we correct for bias in our estimator using the statistical bootstrap method (for a pedagogical introduction see Ref.~\cite{dedeo2013bootstrap}), which also allows us to estimate standard errors about our estimated means.

\subsection*{Calculating focus}

A group's focus is high when individuals strongly concentrate their aggression towards opponents with a particular range of relative rank differences; it is low when aggression is spread across a wider range of individuals. Aggressive events in a group are summarized by the aggression matrix $A$, whose elements $A_{ij}$ count the number of times individual $i$ aggressed against individual $j$.

To define focus we first construct the relative-aggression distribution, $R(\Delta)$, which measures the level of aggression between individuals separated by $\Delta$ steps in relative rank. If we define $P_\Delta$ as the set of all pairs $\{i,j\}$ where $i$ is $\Delta$ ranks above $j$, then $R$ is defined as
\begin{equation}
R(\Delta) = \frac{1}{|P_\Delta|}\sum_{i,j\in P_\Delta} A_{ij},
\end{equation}
where $|P_\Delta|$ is the total amount of aggression by the attackers in the set’s pairs. $R(\Delta)$ is the average amount of aggression directed $\Delta$ rank-steps away. When $\Delta$ is positive, $R(\Delta)$ measures the average aggression directed `down' the hierarchy, from a higher-ranked individual to a lower-ranked individual. 

In other words, $R(\Delta)$ is a measure of the fraction of events that are directed between individuals separated by $\Delta$ steps in relative rank, given the total aggression in the system that could have been directed $\Delta$ steps away. A plot of $R(\Delta)$ as a function of $\Delta$ tells us a great deal about the flows of aggression through the system. 

Focus, $F$, is defined as how ``sharp'' this distribution is:
\begin{equation}
F = 1-\frac{\mathrm{Var}(R)}{N(2N-1)/6},
\end{equation}
where $\mathrm{Var}(R)$ is the $R(\Delta)$-weighted variance of $\Delta$,
\begin{equation}
\mathrm{Var}(R)=\left. \sum_{\Delta=-(N-1)}^{N-1} (\Delta-\bar{\Delta})^2R(\Delta) \middle/ \sum_{\Delta=-(N-1)}^{N-1} R(\Delta) \right.
\end{equation}
and $\bar{\Delta}$ is the $R(\Delta)$-weighted mean of $\Delta$,
\begin{equation}
\bar{\Delta}=\left. \sum_{\Delta=-(N-1)}^{N-1} \Delta R(\Delta) \middle/ \sum_{\Delta=-(N-1)}^{N-1} R(\Delta) \right.
\end{equation}
The normalization term $(2N-1)N/6$ is chosen so that a uniform (flat) distribution of aggression, \emph{i.e.}, ``rank ignorant'', gives a focus of zero. If focusing is very strong---\emph{e.g.}, if all individuals direct their aggression towards the individual two ranks down from them in the hierarchy, $F$ is 1. As aggression is more evenly distributed, $F$ decreases. In the case that aggression is completely uniform across all ranks, then the normalization is chosen such that $F$ will be precisely zero. (In rare cases, where the aggression is ``overdispersed'', it is possible to have negative focus.)
 
\subsection*{Position of focused aggression} 

If rank information is present and is used, and we can detect this via focus, then knowing the position of the peak of aggression gives us information about the specific relative rank-based aggression pattern that individuals are using. For example, individuals with focused aggression could direct most of their aggression towards those that are ranked directly beneath themselves in the hierarchy. Alternatively, individuals could focus their aggression on the very lowest ranked individuals in the group. These two cases could result in similar levels of focus in aggression, but could be differentiated from each other by differences in their position values. In the first case, position would be closer to each individual's own rank (and closer to 0) while in the second case, position would move towards 1 as aggression is directed at individuals many ranks distant from an individual's own rank.  

We define the position of focused aggression as the average of the distribution of normalized aggression for each social group; \emph{i.e.}, for each individual, we compute the probability that the individual's aggression is directed at an individual rank $\Delta$ away, $P_i(\Delta)$, and then average these probabilities over all individuals, formally,
\begin{equation}
P = \sum_{i=1}^N \sum_{\Delta\in O(i)} \Delta P_i(\Delta),
\end{equation}
where $O(i)$ is the list of relative ranks available to individual $i$; higher-ranked individuals have more relative ranks available downward (positive $\Delta$), while lower-ranked individuals have more available up.

The $P$ measure accounts for the effects of both individual aggression levels and the number of potential aggressive targets as a function of rank, and allows us to capture the extent to which decision-making on the individual level is sensitive to relative rank position. 

\subsection*{Modelling the structural rules of dominance hierarchies}

We compare the focus and position values quantified from observed empirical data with those generated from (1) a set of permutation-based reference models, where we use a complex edge rewiring procedure that reproduces the basic hierarchical structure found in each empirical group, and (2) a set of generative models that use an agent-based framework to simulate social systems with winner and/or loser effects. This approach allows us to determine minimal models for group-level aggression patterns used in a particular group. 

The simplest rank-based rule we considered is the  ``downward heuristic'', where individuals aggress only against those ranked below themselves. We used this  rule to recreate aggression networks for each group, and compared it with the  observed aggression, using an ensemble of agent-based model simulations to create a set of reference aggression networks. These models preserve some of the basic structure we observe in data, such as the overall aggressive dispositions of each individual, while potentially  permuting other aspects.

We used these reference aggression networks to determine which values of focus and position we should expect to be generated if animals in the group were only using the downward heuristic. We use each individual's rank, calculated from the data, and then allow individuals to aggress as much as they do in the empirical data, but, in the simplest case, with a uniform random preference for aggression against only the individuals ranked below themselves in the hierarchy. This process is consistent with best practice recommendations for animal social network permutation, which supports event-level permutations of social interactions rather than relationship strengths~\cite{farine2017networknulls}. 

Formally, given the aggression matrix $A_{ij}$, and the ranks $r_i$. The first-ranked individual, $k$, has $r_k$ equal to one, and $r_i > r_j$ indicates that $i$ is lower ranked than $j$. Then, for each individual $i$, the row $A_{ij}$ is then mapped to $A^\prime_{ij}$ where
\begin{equation}
A^\prime_{ij} = \frac{\sum_{j=1}^N A_{ij}}{N-r_i+1}\delta_{r_j>r_i},
\end{equation}
and $\delta_{r_j>r_i}$ is equal to one when the subscript is true (\emph{i.e.}, when then $j$ is lower ranked than $i$) and zero otherwise. This mapping takes the total aggression by individual $i$ and distributes it equally towards all lower-ranked individuals.

Empirical systems may be somewhat noisy and may not follow a pure downward heuristic (\textit{e.g.}, due to mistakes in directing aggression, occasional opportunism in attacking a higher-ranked individual, or some level of stochasticity in directing aggression). At the extreme, individuals may direct aggression based only on their own levels of aggression, in complete disregard for rank differences. 

To account for this, we introduced the possibility of randomness in aggression direction; mathematically, we allow for an $\epsilon$ probability that the individual simply directs aggression at a random individual,
\begin{equation}
A^\prime_{ij;\epsilon} = (1-\epsilon)A^\prime_{ij} + \frac{\epsilon\delta_{i\neq j}}{N-1}.
\end{equation}
We conducted a parameter sweep of the downward aggression heuristic in $\epsilon$, gradually increasing the amount of randomly-directed aggression from $\epsilon$ equal to zero (perfect downward aggression) to unity (completely randomly-directed aggression based only on individual aggressiveness), then examined how increasing randomness affected focus and position values. This process allowed us to simulate aggression along a continuum, from perfect use of basic rank information to completely random behavior dictated solely by each individual's own levels of observed aggressiveness.


\section*{Social dominance pattern assignment}
Our reference aggression networks generated by the downward heuristic serve as randomized reference models~\citep{Gauvin2018} to which we can compare the observed datasets, and as a form of null model for the downward heuristic: we fail to reject the downward heuristic as a plausible generating rule of focus and position in the observed datasets if observed focus and position values fall within the range that can be produced by our refenrence model datasets. For observed groups that fall outside of the region that could be generated by the downward heuristic, we categorize these groups into social dominance patterns other than the basic downward heuristic.

We categorized groups into thee main social dominance pattern types: downward heuristic, close competitors, and bullying. We ran a suite of agent-based models of aggression under the downward heuristic pattern, scanning across values of $\epsilon$ from zero (perfect use of categorical rank information) to one (completely random behavior based only on individual aggressiveness). This enabled us to delineate the focus and position parameter space in which these summary measures are  consistent with those produced by the downward heuristic. We drew a polygon around the space traced out by different values of $\epsilon$, using the extremes of error bars to set the edges of the polygon ($95\%$ CI, Fig.~\ref{strategy.exs}). Observed data that intersected this downward heuristic polygon were scored as consistent with that model if any of the error bars for the observed data overlapped with the polygon (Fig.~\ref{strategy.exs}a). Fig.~\ref{strategy.exs} provides an example of these assignments. This procedure, fit to aspects of each of the observed social groups, allowed us to discriminate between social dominance patterns on a case by case basis rather than using a generalized rule for all focus and position values (as a result, the same observed value of focus and position may be categorized as ``close competitors'' in  some groups and ``downward heuristic'' in others, see Fig.~\ref{FxP}).

We defined the close competitors social dominance pattern as having a lower position value than that produced by the downward heuristic model (\emph{i.e.} aggression concentrated on opponents ranked just below themselves, Fig.~\ref{strategy.exs}b) and bullying as having a higher position value than the modelled data (\emph{i.e.} aggression concentrated on opponents ranked far below themselves, Fig.~\ref{strategy.exs}c). Some groups had undefined social dominance patterns with focus values lower than those expected in fully random systems.  

\subsection*{Phylogenetic analyses}
We used the {\tt R} package \textit{taxize}~\citep{Rtaxizepaper,Rtaxizemanual} to check all species names, assign them to order, and plot the phylogenetic relationships among the 23 orders. To test for an effect of relatedness across taxa, we took the social dominance patterns for each group in our dataset and randomly re-allocated all strategies (without replacement) so that each group had a new randomly-assigned pattern. For all species, we then summarized the occurrence of social dominance patterns for each of the 23 orders for each of 1000 randomization runs. We compared the frequency with which each of the three main social dominance patterns was observed in each order to the frequencies expected if social dominance is randomly assigned. If the observed frequency falls within this expected distribution, we concluded that we have no evidence that relatedness among taxa (at the order level) affects which groups use each social dominance pattern.

\subsection*{General analyses and code availability}
All final analyses were run in {\tt R}~\citep{Rlang}. We used {\tt R} packages gplots~\citep{Warnes2020gplots} and ggtree~\citep{Yu2017,Yu2018} to plot Figure~\ref{StratsxOrder} and {\tt R} package ggridges~\citep{Wilke2020ggridges} to plot Figures~\ref{ridgeplot_order},~\ref{ridgeplot_grptype}, and~\ref{ridgeplot_grpsize}. Code to enable running the analyses is contained in the {\tt R} package \textit{domstruc} (\url{https://github.com/danm0nster/domstruc}) and all code for analyses and visualizations will be made available on github on publication of the paper.

\section*{Acknowledgements}

EAH was supported by a postdoctoral fellowship from the ASU-SFI Center for Biosocial Complex Systems, with additional funding from the Santa Fe Institute. DM was funded in part by Independent Research Fund Denmark (grant no.\ 7089-00017B), Aarhus University Research Foundation, and The Interacting Minds Centre, and gratefully acknowledges the hospitality of the Santa Fe Institute during a sabbatical visit. This research was additionally supported by Army Research Office Grant \#W911NF1710502. EAH would like to thank Joshua Garland, Brendan Tracey, Vanessa Ferdinand, and Andy Rominger for many helpful discussions which have improved the drafts. 

\clearpage
\bibliographystyle{naturemag}
\bibliography{MASTERBIBall,MendeleyRefs,inprep}

\clearpage
\newpage

%
%

\newpage
\clearpage
\pagenumbering{arabic}
\renewcommand{\thepage}{\arabic{page}}
\renewcommand{\thesection}{SI \arabic{section}}   

\section*{Supplementary Information}
\beginsupplement
\hyphenation{Page-Rank} 

\section{Setting the regularization term for the measurement of PageRank in animal conflict} \label{SI-epsilon}
\setcounter{figure}{0}
\renewcommand{\thefigure}{S\arabic{section}.\arabic{figure}} 
\setcounter{table}{0}
\renewcommand{\thetable}{S\arabic{section}.\arabic{table}} 

A basic step in the calculation of focus and average peak position is the estimation of the transition matrix, $T_{ij}$, a collection of probabilities, from the data. The ``naive'' way to estimate a probability of an event occurring from a finite number of observations is
\begin{equation}
\tilde{p} = \frac{n_i}{N}.
\end{equation}
While attractive in its simplicity, this estimator has a number of problems (see Ref.~\cite{jaynes2003probability}); a Bayesian analysis leads to the correction
\begin{equation}
\hat{p} = \frac{n_i+\epsilon}{N+m\epsilon},
\label{bayes}
\end{equation}
where $m$ is the number of event types, and $\epsilon$ a regularization parameter (sometimes called a ``teleportation term''). When $\epsilon$ is equal to unity, we have Laplace's rule; more generally, we can think of $\epsilon$ as parametrising a Dirichlet distribution that serves as the prior for the possible values of the underlying probabilities $p$~\cite{wolpert1995estimating, wolpert2013estimating}.

In the case we have here, $T_{ij}$ is the estimate of the probability that $i$ against $j$; by stipulation, the individual $i$ can not aggress against itself. We can then adapt equation~\ref{bayes} to the estimate of the probability distributions in the matrix $T_{ij}$.

How do we choose $\epsilon$? A natural way to do so is to learn $\epsilon$ from the data itself; we do so here using $k$-fold cross validation, with $k$ set to five. For each dataset, in other words, we compute the probabilities $T_{ij}$, for some particular choice of $\epsilon$, based on a randomly chosen sample of only $4/5$ of the data. We then compute the log-probability per data-point of the remaining ``held out'' $1/5$ of the data, $n_{ij}$, using those estimated $T_{ij}$s,
\begin{equation}
L(\epsilon) = \frac{1}{N_h}\sum_{i,j=1}^N n_{ij}\log{T_{ij}(\epsilon)},
\label{clever}
\end{equation}
where $N_h$ is the number of observations in the held-out set (\emph{i.e.}, $1/5$ of the total number of observations). In words, $L(\epsilon)$ is how well that particular choice of $\epsilon$ ``predicts'' the held-out data; the optimal choice of $\epsilon$ is that which best predicts.

\begin{figure}[h]
\caption{Determining optimal $\epsilon$ through $k$-fold cross validation; an example of equation~\ref{clever} applied to the data of Ref.~\cite{allee1954dominance}. An $\epsilon$ value of approximately $0.3$, in this case, best predicts held-out data, but a range of $\epsilon$ values between $0.1$ and $1$ perform similarly well.}
\label{allee_test}
\includegraphics[width=.5\textwidth]{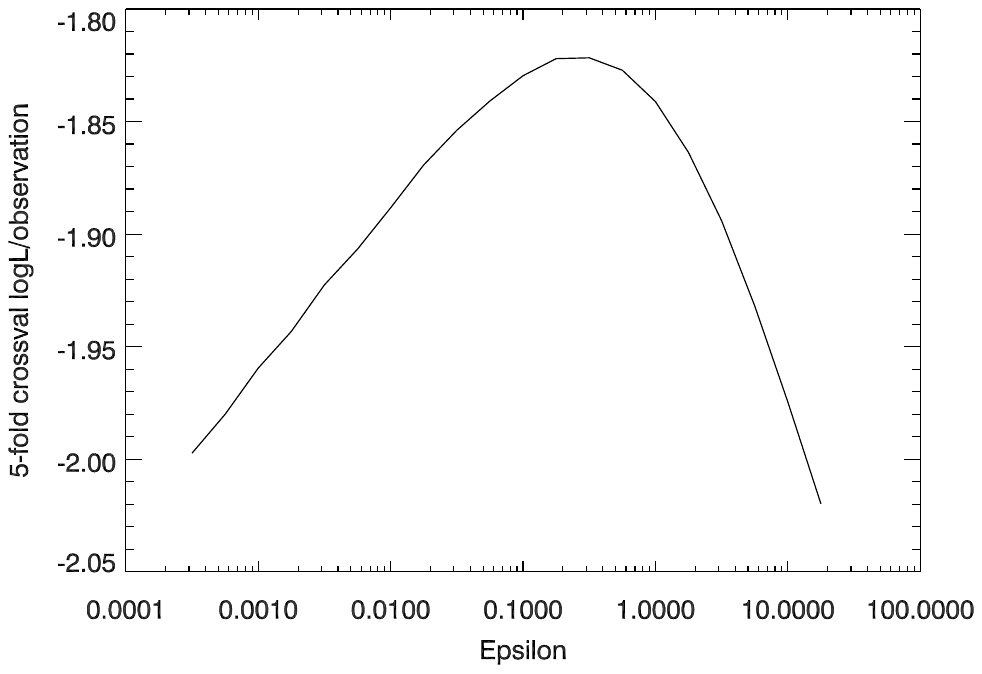}
\end{figure}

\begin{figure}[h]
\caption{A scatter plot of optimal epsilons found using equation~\ref{clever}, as a function of total number of observations (left), and total number of individuals in the data (right). The optimal value shows no strong trends with either variable; the average optimal value for $\epsilon$ is $0.694$ and we use this for simplicity in the calculations in the main text.}
\label{scatter}
\includegraphics[width=.495\textwidth]{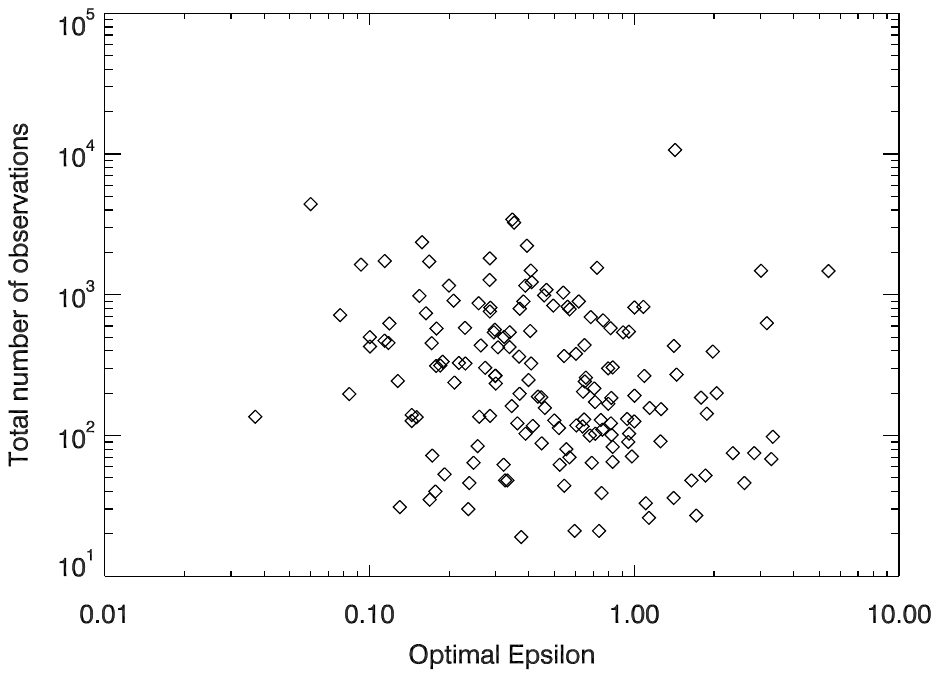} 
\includegraphics[width=.495\textwidth]{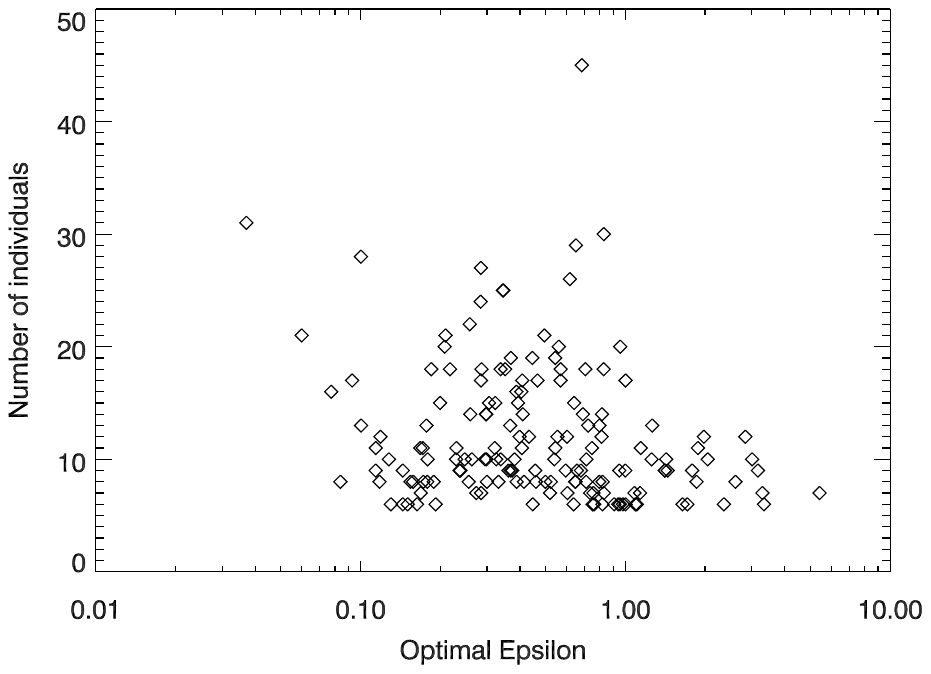}
\end{figure}

We repeat this process many times, choosing a different hold-out set each time, to get an estimate of the average log-probability of the held-out data. We then choose $\epsilon$ to maximize this average of $L(\epsilon)$. Fig.~\ref{allee_test} shows an example of this process for the data of Ref.~\cite{allee1954dominance}. The peak of this function allows us to pick the optimal epsilon to be around $0.3$ for this dataset, although values between $0.2$ and $0.6$ are largely indistinguishable. Fig.~\ref{scatter} shows a scatter plot of the $L$-maximizing $\epsilon$ for all 161 aggression matrices in our data, as a function of both total number of observations, and number of individuals. 

We find that most matrices have optimal values of $\epsilon$ between $0.1$ and $1.0$, and that there is no strong correlation between optimal $\epsilon$ and system size or total number of observations. The average value of epsilon across all datasets is $0.694$.

Little hinges on the exact value of $\epsilon$; indeed, using the average value in place of the optimal choice for any particular dataset leads to an average (absolute value) shift in the focus measure of only $0.027$, and in the average peak position of only $0.017$; over our data, the two choices have a Pearson correlation of $0.95$ (Focus) and $0.97$ (Average Peak Position). Since finding the optimal $\epsilon$ is computationally intensive, and since the final results are largely insensitive to this choice, we suggest the average value, $0.694$, is appropriate for ordinary use, and (for simplicity) we present our analyses here using this choice.

\newpage
\section{Observed summary data} \label{SI-data}
\setcounter{figure}{0}
\renewcommand{\thefigure}{S\arabic{section}.\arabic{figure}} 
\setcounter{table}{0}
\renewcommand{\thetable}{S\arabic{section}.\arabic{table}} 

\begin{longtable}{llrrl}
\caption{Observed social dominance pattern, focus, and position values for each empirical group in our dataset (focus and position values rounded). Social dominance patterns: downward heuristic (DH), close competitors (CC), bullying (BL), and undefined (UN). Records sorted by species name then by file name.}
\label{tab:data}\\
\hline
Species & Pattern & Focus & Position & File name \\ 
\hline 
\endfirsthead 
\hline
Species & Pattern & Focus & Position & File name \\ 
\hline 
\endhead 
\hline \multicolumn{4}{r}{\textit{Continued on next page}} \\
\endfoot
\hline 
\endlastfoot
Addax nasomaculatus & CC & 0.78 & 0.34 & Reason1988-1.csv \\ 
  Anas acuta & DH & 0.69 & 0.36 & Poisbleau2005-1b.csv \\ 
  Anas platyrhynchos & DH & 0.74 & 0.46 & Poisbleau2005-1a.csv \\ 
  Anas platyrhynchos & BL & 0.45 & 0.38 & Poisbleau2006-2a.csv \\ 
  Anolis aeneus & CC & 0.86 & 0.30 & Stamps1978.csv \\ 
  Antilocapra americana & DH & 0.60 & 0.29 & Bromley1991-1.csv \\ 
  Antilocapra americana & DH & 0.71 & 0.37 & Fairbanks1994-5.csv \\ 
  Antilocapra americana & DH & 0.41 & 0.22 & Fairbanks1994-6.csv \\ 
  Antilocapra americana & BL & 0.66 & 0.58 & Fairbanks1994-7.csv \\ 
  Bison bison & DH & 0.44 & 0.30 & Lott1979-1.csv \\ 
  Bison bison & DH & 0.64 & 0.34 & Lott1987-1.csv \\ 
  Bison bison & DH & 0.24 & 0.28 & Rutberg1986-2.csv \\ 
  Branta bernicla & DH & 0.55 & 0.31 & Poisbleau2006-2b.csv \\ 
  Callosciurus erythraeus & DH & 0.62 & 0.37 & Tamura1988-1.csv \\ 
  Canis lupus & DH & 0.55 & 0.36 & Cafazzo2010-5.csv \\ 
  Cercopithecus mitis & DH & 0.78 & 0.38 & Payne2003-1.csv \\ 
  Cervus elaphus & DH & 0.17 & 0.19 & Appleby1983-1.csv \\ 
  Chlorocebus aethiops & DH & 0.71 & 0.42 & Isbell1998-A.csv \\ 
  Chlorocebus aethiops & DH & 0.75 & 0.38 & Struhsaker1967-6.csv \\ 
  Chlorocebus aethiops & DH & 0.60 & 0.34 & Struhsaker1967-7.csv \\ 
  Colobus polykomos & DH & 0.48 & 0.42 & Korstjens2002-1.csv \\ 
  Columba livia & UN & -0.05 & 0.15 & Masure1934-3.csv \\ 
  Corvus monedula & DH & 0.50 & 0.34 & Roell1978-11.csv \\ 
  Corvus monedula & DH & 0.27 & 0.23 & Tamm1977-1b.csv \\ 
  Crocuta crocuta & DH & 0.56 & 0.45 & Frank1986-1.csv \\ 
  Crocuta crocuta & DH & 0.72 & 0.37 & Holekamp1991-1.csv \\ 
  Crocuta crocuta & DH & 0.68 & 0.50 & Holekamp1993-1a.csv \\ 
  Crocuta crocuta & DH & 0.74 & 0.29 & Jenks1995-3.csv \\ 
  Crocuta crocuta & BL & 0.73 & 0.56 & Tilson1984-1.csv \\ 
  Cyanocitta cristata & DH & 0.61 & 0.36 & Tarvin1997-5.csv \\ 
  Dinoponera quadriceps & DH & 0.18 & 0.24 & Monnin1999-1.csv \\ 
  Eledone moschata & DH & 0.48 & 0.20 & Mather1985.csv \\ 
  Emys orbicularis & DH & 0.25 & 0.26 & Rovero1999-2a.csv \\ 
  Emys orbicularis & DH & 0.12 & 0.18 & Rovero1999-2b.csv \\ 
  Equus caballus & CC & 0.74 & 0.39 & CluttonBrock1976-3.csv \\ 
  Equus caballus & CC & 0.78 & 0.35 & Ellard1989-3.csv \\ 
  Equus caballus & BL & 0.71 & 0.44 & Heitor2006-3.csv \\ 
  Equus caballus & DH & 0.49 & 0.29 & Heitor2010-2.csv \\ 
  Equus caballus & CC & 0.65 & 0.38 & Wells1979-3.csv \\ 
  Erythrocebus patas & BL & 0.46 & 0.58 & Isbell1998-B.csv \\ 
  Felis catus & DH & 0.66 & 0.45 & Bonanni2007-2.csv \\ 
  Felis catus & BL & 0.50 & 0.54 & Natoli1991-2.csv \\ 
  Fringilla coelebs & CC & 0.86 & 0.37 & Marler1955b.csv \\ 
  Galaxias argenteus & DH & 0.53 & 0.37 & David2003-2a.csv \\ 
  Galaxias argenteus & DH & 0.53 & 0.44 & David2003-2b.csv \\ 
  Gorilla beringei & DH & 0.37 & 0.24 & Robbins2008-2.csv \\ 
  Gorilla gorilla & DH & 0.24 & 0.33 & Scott1999-2b.csv \\ 
  Gorilla gorilla & DH & 0.13 & 0.19 & Scott1999-2c.csv \\ 
  Haemorhous mexicanus & DH & 0.72 & 0.40 & Thompson1960-JJ59A.csv \\ 
  Haemorhous mexicanus & DH & 0.71 & 0.41 & Thompson1960-JJ59B.csv \\ 
  Haemorhous mexicanus & DH & 0.74 & 0.39 & Thompson1960-JJ59C.csv \\ 
  Haemorhous mexicanus & DH & 0.52 & 0.31 & Thompson1960-ND54.csv \\ 
  Haemorhous mexicanus & DH & 0.37 & 0.21 & Thompson1960-ND57.csv \\ 
  Haliaeetus albicilla & DH & 0.36 & 0.29 & Kolodziejczyk2005-1.csv \\ 
  Junco hyemalis & DH & 0.24 & 0.31 & Yasukawa1983-1a.csv \\ 
  Junco hyemalis & DH & 0.27 & 0.30 & Yasukawa1983-1b.csv \\ 
  Junco hyemalis & DH & 0.08 & 0.22 & Yasukawa1983-2a.csv \\ 
  Junco hyemalis & DH & -0.05 & 0.08 & Yasukawa1983-2b.csv \\ 
  Junco hyemalis & DH & 0.64 & 0.37 & Yasukawa1983-5a.csv \\ 
  Junco hyemalis & DH & 0.58 & 0.29 & Yasukawa1983-5b.csv \\ 
  Junco hyemalis & BL & 0.53 & 0.43 & Yasukawa1983-6a.csv \\ 
  Junco hyemalis & DH & 0.57 & 0.24 & Yasukawa1983-6b.csv \\ 
  Lampropholis guichenoti & DH & 0.35 & 0.22 & Torr1996-1.csv \\ 
  Leptothorax sp. & DH & 0.35 & 0.21 & Ortius1995-2a.csv \\ 
  Loxodonta africana & CC & 0.77 & 0.32 & Archie2006-A.csv \\ 
  Loxodonta africana & DH & 0.67 & 0.42 & Archie2006-AA.csv \\ 
  Loxodonta africana & DH & 0.50 & 0.36 & Archie2006-CB.csv \\ 
  Loxodonta africana & DH & 0.68 & 0.37 & Archie2006-FB.csv \\ 
  Loxodonta africana & DH & 0.57 & 0.35 & Archie2006-GB.csv \\ 
  Loxodonta africana & DH & 0.52 & 0.51 & Archie2006-JA.csv \\ 
  Loxodonta africana & DH & 0.74 & 0.40 & Archie2006-OA.csv \\ 
  Loxodonta africana & BL & 0.91 & 0.55 & Archie2006-PC.csv \\ 
  Loxodonta africana & DH & 0.77 & 0.36 & Archie2006-SI.csv \\ 
  Loxodonta africana & DH & 0.50 & 0.42 & Wittemeyer2007-1.csv \\ 
  Macaca arctoides & DH & 0.40 & 0.41 & Richter2009-1.csv \\ 
  Macaca fascicularis & DH & 0.36 & 0.33 & deWaal1977-1.csv \\ 
  Macaca fascicularis & BL & 0.49 & 0.37 & deWaal1977-2.csv \\ 
  Macaca fascicularis & DH & 0.65 & 0.42 & Sterck1997-4KB.csv \\ 
  Macaca mulatta & DH & 0.71 & 0.44 & deWaal1985-1.csv \\ 
  Macaca mulatta & DH & 0.13 & 0.18 & Varley1966-1.csv \\ 
  Macaca thibetana & DH & 0.60 & 0.41 & Berman2004-B.csv \\ 
  Mandrillus sphinx & CC & 0.64 & 0.30 & Setchell2005-3.csv \\ 
  Mandrillus sphinx & DH & 0.58 & 0.36 & Setchell2005-4.csv \\ 
  Mareca penelope & DH & 0.53 & 0.39 & Poisbleau2005-1c.csv \\ 
  Melanochromis auratus & DH & 0.27 & 0.26 & Nelissen1985-2.csv \\ 
  Meles meles & DH & 0.60 & 0.46 & Hewitt2009-P2005.csv \\ 
  Meles meles & DH & 0.48 & 0.37 & Hewitt2009-PO2004.csv \\ 
  Meles meles & DH & 0.50 & 0.42 & Hewitt2009-SH1995.csv \\ 
  Mustelus canis & DH & 0.73 & 0.42 & Allee1954-3.csv \\ 
  Myiopsitta monachus & CC & 0.65 & 0.30 & monkparakeet.g1Q2to4.csv \\ 
  Myiopsitta monachus & CC & 0.65 & 0.27 & monkparakeet.g2.Q2to4.noNBB.csv \\ 
  Neoponera villosa & CC & 0.75 & 0.36 & Trunzer1999-1.csv \\ 
  Notamacropus parryi & DH & 0.54 & 0.30 & Kaufmann1974-7.csv \\ 
  Nymphicus hollandicus & DH & 0.31 & 0.29 & Seibert2001-3.csv \\ 
  Odocoileus hemionus & DH & 0.69 & 0.37 & Koutnik1981-3.csv \\ 
  Odocoileus virginianus & DH & 0.77 & 0.44 & Collias1950-1.csv \\ 
  Oncorhynchus masou & CC & 0.83 & 0.39 & Nakano1994-1.csv \\ 
  Oreamnos americanus & DH & 0.73 & 0.43 & Cote2000-4.csv \\ 
  Oreamnos americanus & DH & 0.56 & 0.41 & Fournier1995-3.csv \\ 
  Osphranter rufus & DH & 0.53 & 0.39 & Russell1970-2a.csv \\ 
  Osphranter rufus & DH & 0.57 & 0.30 & Russell1970-2b.csv \\ 
  Ovis canadensis & DH & 0.63 & 0.35 & Hass1991-2.csv \\ 
  Ovis canadensis & CC & 0.48 & 0.17 & Zine2000-1.csv \\ 
  Pachycondyla sp. & DH & 0.69 & 0.42 & Ito1993-2a.csv \\ 
  Pachycondyla sp. & DH & 0.62 & 0.40 & Ito1993-3a.csv \\ 
  Pachycondyla sp. & DH & 0.70 & 0.31 & Ito1993-4a.csv \\ 
  Pan paniscus & BL & 0.27 & 0.42 & Paoli2006-2.csv \\ 
  Pan paniscus & BL & 0.65 & 0.54 & Vervaecke2000-2.csv \\ 
  Pan troglodytes & DH & 0.53 & 0.42 & Murray2007-3.csv \\ 
  Pan troglodytes & CC & 0.70 & 0.36 & Wittig2003-1.csv \\ 
  Papio cynocephalus & DH & 0.19 & 0.13 & Cheney1977-B.csv \\ 
  Papio cynocephalus & DH & 0.85 & 0.32 & Hausfater1975-10.csv \\ 
  Papio cynocephalus & CC & 0.81 & 0.23 & Hausfater1975-11.csv \\ 
  Papio cynocephalus & CC & 0.88 & 0.29 & Hausfater1975-6.csv \\ 
  Papio cynocephalus & CC & 0.80 & 0.38 & Hausfater1982-1.csv \\ 
  Papio cynocephalus & CC & 0.73 & 0.38 & Hausfater1982-2.csv \\ 
  Papio cynocephalus & DH & 0.61 & 0.40 & Lee1979-1.csv \\ 
  Papio cynocephalus & DH & 0.52 & 0.41 & Lee1979-2.csv \\ 
  Papio cynocephalus & DH & 0.64 & 0.31 & McMahan1984-1.csv \\ 
  Papio cynocephalus & CC & 0.78 & 0.37 & Samuels1987-2.csv \\ 
  Parahyaena brunnea & DH & 0.75 & 0.43 & Owens1996-1.csv \\ 
  Passer domesticus & BL & 0.41 & 0.36 & Moller1987-1.csv \\ 
  Passer domesticus & BL & 0.70 & 0.44 & Moller1987-2.csv \\ 
  Passer domesticus & DH & 0.73 & 0.48 & Moller1987-3.csv \\ 
  Passer domesticus & DH & 0.13 & 0.25 & Solberg1997-1.csv \\ 
  Passer domesticus & DH & 0.34 & 0.21 & Solberg1997-2.csv \\ 
  Passer domesticus & CC & 0.67 & 0.01 & Solberg1997-3.csv \\ 
  Phasianus colchicus & DH & 0.88 & 0.42 & Collias1951-3.csv \\ 
  Phasianus colchicus & DH & 0.74 & 0.26 & Collias1951-4.csv \\ 
  Phasianus colchicus & BL & 0.53 & 0.46 & Collias1951-5.csv \\ 
  Phoca vitulina & DH & 0.56 & 0.40 & Sullivan1982-4.csv \\ 
  Poecile atricapilla & DH & 0.62 & 0.41 & Hartzler1970-1.csv \\ 
  Poecile atricapilla & CC & 0.85 & 0.39 & Smith1976-1.csv \\ 
  Poecile montanus & DH & 0.57 & 0.35 & Lahti1994-A.csv \\ 
  Poecile montanus & DH & 0.42 & 0.34 & Lahti1994-C.csv \\ 
  Polistes canadensis & DH & 0.37 & 0.26 & West-Eberhard1986-5.csv \\ 
  Potamonautes perlatus & DH & 0.54 & 0.38 & Somers1998-2a.csv \\ 
  Potamonautes perlatus & DH & 0.53 & 0.37 & Somers1998-2b.csv \\ 
  Potamonautes perlatus & DH & 0.79 & 0.45 & Somers1998-2c.csv \\ 
  Protomognathus americanus & DH & 0.80 & 0.28 & Blatrix2004-2.csv \\ 
  Protomognathus americanus & DH & 0.72 & 0.39 & Blatrix2004-3.csv \\ 
  Protomognathus americanus & DH & 0.74 & 0.38 & Blatrix2004-5.csv \\ 
  Protomognathus americanus & DH & 0.72 & 0.34 & Blatrix2004-6.csv \\ 
  Quiscalus major & DH & 0.65 & 0.40 & Post1992-1.csv \\ 
  Rangifer tarandus & DH & 0.61 & 0.34 & Barette1986.csv \\ 
  Rangifer tarandus & DH & 0.51 & 0.34 & Hirotani1994-1.csv \\ 
  Salvelinus leucomaenis & DH & 0.60 & 0.36 & Nakano1995-2.csv \\ 
  Sapajus apella & DH & 0.79 & 0.47 & Izar2006-2.csv \\ 
  Sauromalus ater & DH & 0.30 & 0.20 & Prieto1978-1.csv \\ 
  Sciurus aberti & DH & 0.73 & 0.34 & Farentinos1972-D.csv \\ 
  Semnopithecus entellus & DH & 0.60 & 0.31 & Lu2008-1c.2.csv \\ 
  Serinus canaria & DH & 0.04 & 0.10 & Shoemaker1939-1.csv \\ 
  Sphyrna tiburo & DH & 0.44 & 0.27 & Myrberg1974-17.csv \\ 
  Stegastes partitus & CC & 0.82 & 0.30 & Myrberg1972-3.csv \\ 
  Streptopelia risoria & DH & 0.11 & 0.15 & Bennett1939-2.csv \\ 
  Tadorna tadorna & DH & 0.63 & 0.38 & Patterson1977-6.csv \\ 
  Trachypithecus phayrei & DH & 0.24 & 0.29 & Koenig2004-1b.csv \\ 
  Tragelaphus angasii & DH & 0.75 & 0.34 & Collias1950-2.csv \\ 
  Tropheus moorii & DH & 0.63 & 0.39 & Kohda1991-2.csv \\ 
  Vollenhovia nipponica & DH & 0.08 & 0.10 & Satoh-C.csv \\ 
  Vollenhovia nipponica & DH & 0.01 & 0.11 & Satoh-D.csv \\ 
  Vollenhovia nipponica & DH & 0.38 & 0.25 & Satoh-E.csv \\ 
  Xerus rutilus & DH & 0.70 & 0.41 & Oshea1976-1.csv \\ 
  Zonotrichia leucophrys & DH & 0.13 & 0.10 & Parsons1980-2a.csv \\ 
  Zonotrichia leucophrys & BL & 0.60 & 0.41 & Slotow1993-1.csv \\ 
  Zonotrichia querula & BL & 0.72 & 0.51 & Watt1986-1a.csv \\ 
  Zonotrichia querula & DH & 0.72 & 0.41 & Watt1986-1b.csv \\ 
  Zonotrichia querula & DH & 0.69 & 0.38 & Watt1986-1c.csv \\ 
  Zonotrichia querula & DH & 0.37 & 0.38 & Watt1986-1d.csv \\ 
  Zonotrichia querula & DH & 0.59 & 0.41 & Watt1986-1e.csv \\ 
  Zosterops lateralis & DH & 0.32 & 0.18 & Kikkawa1980-1.csv \\ 
  Zosterops lateralis & DH & 0.21 & 0.15 & Williams1972-1.csv \\ 
  \hline
\end{longtable}

\newpage
\section{Characteristics of structured aggression} \label{SI-strucagg}
\setcounter{figure}{0}
\renewcommand{\thefigure}{S\arabic{section}.\arabic{figure}} 
\setcounter{table}{0}
\renewcommand{\thetable}{S\arabic{section}.\arabic{table}} 

Most of the animal social groups in the empirical dataset had well-structured dominance hierarchies. Groups generally had real focus values consistent with low levels of randomly-directed aggression: $46\%$ of groups ($N=79$) were most similar to modelled data with $10\%$ or less randomly-directed events ($\leq0.1$); $63\%$ of groups ($N=109$) were most similar to modelled data with $20\%$ or less randomly-directed events ($\leq0.2$) and $90\%$ of groups ($N=155$) was most similar to modeled data with $60\%$ or less randomly-directed events. 

Only 14 groups ($8\%$) had focus values most similar to modelled data with $80\%$ or greater randomly-directed events ($\geq0.8$); of these, none were categorized as a close competitor or bullying social dominance type. 

Only 7 groups ($4\%$) had focus values closest to modelled data with totally random aggression, which corresponds roughly to previous results with this dataset which found over-representation of transitive configurations, an indication of structured hierarchies, in all but $3\%$ of groups~\citep{shizuka2015domhier}.

\newpage
\section{Ruling out factors affecting observed social dominance patterns} \label{SI-factors}

\setcounter{figure}{0}
\renewcommand{\thefigure}{S\arabic{section}.\arabic{figure}} 
\setcounter{table}{0}
\renewcommand{\thetable}{S\arabic{section}.\arabic{table}} 

Factors other than phylogenetic relatedness could affect which social dominance pattern a social group used. Here we investigate whether the occurrence of social dominance patterns can be explained by either the conditions under which groups were observed (if data were collected from a natural population or one held in captivity) or the number of individuals in the group (for example, if only small groups showed evidence of a certain social dominance pattern).

We use the same randomized data as in the main text, where we tested whether randomizing social dominance types across orders would differ from observed occurrences. Here, we summarize by (1) observation conditions and (2) group size. If the observed pattern occurrences fall within the distributions of the randomized strategies, then we can conclude that we do not have any evidence that the factors explain the observed distributions.

\begin{figure} [!htb]
\centering
\caption{The conditions under which groups were observed had little to no relationship with social dominance patterns. For groups observed in natural or captive conditions, the observed number of groups with each social dominance pattern (solid vertical lines) overlaps with the number of groups with each pattern when patterns are randomly allocated (shaded areas, density estimates) for each of the three main social dominance patterns (a) downward heuristic, (b) close competitors, and (c) bullying.} 
\label{ridgeplot_grptype}
\includegraphics[width=0.9\textwidth]{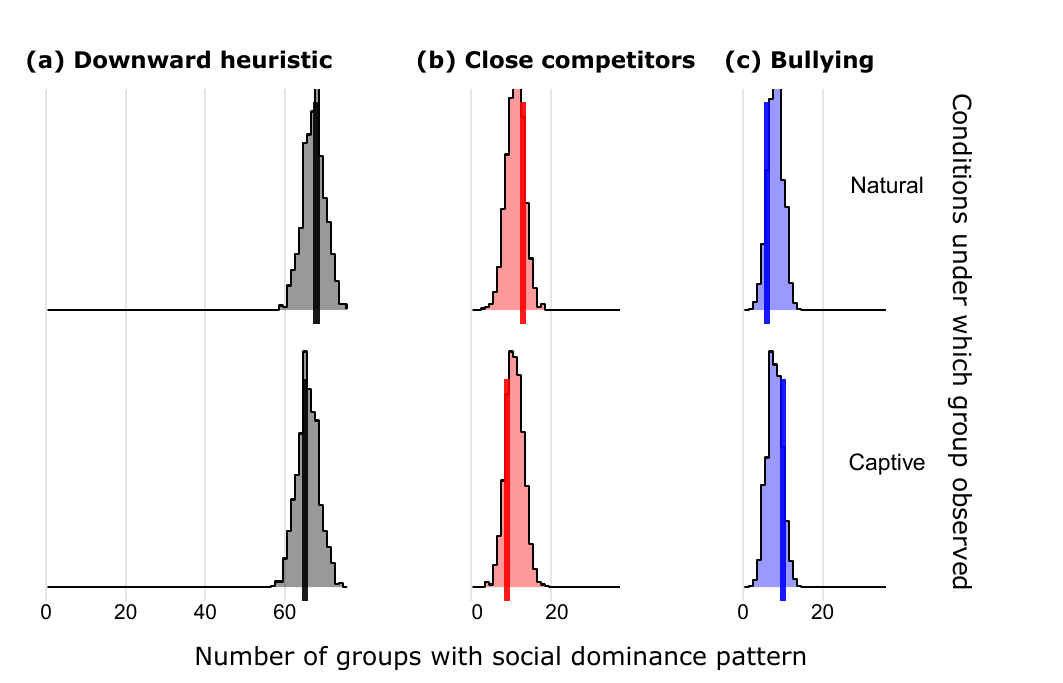}
\end{figure}

\begin{figure} [!htb]
\centering
\caption{The size of observed social groups had little to no relationship with social dominance patterns. Across different group sizes, the number of individuals in the group did not consistently explain the occurrence of each social dominance pattern (solid vertical lines): the observed number of groups overlaps with the number of groups with each pattern when patterns are randomly allocated (shaded areas, density estimates) for each of the three main social dominance patterns (a) downward heuristic, (b) close competitors, and (c) bullying.} 
\label{ridgeplot_grpsize}
\includegraphics[width=1.0\textwidth]{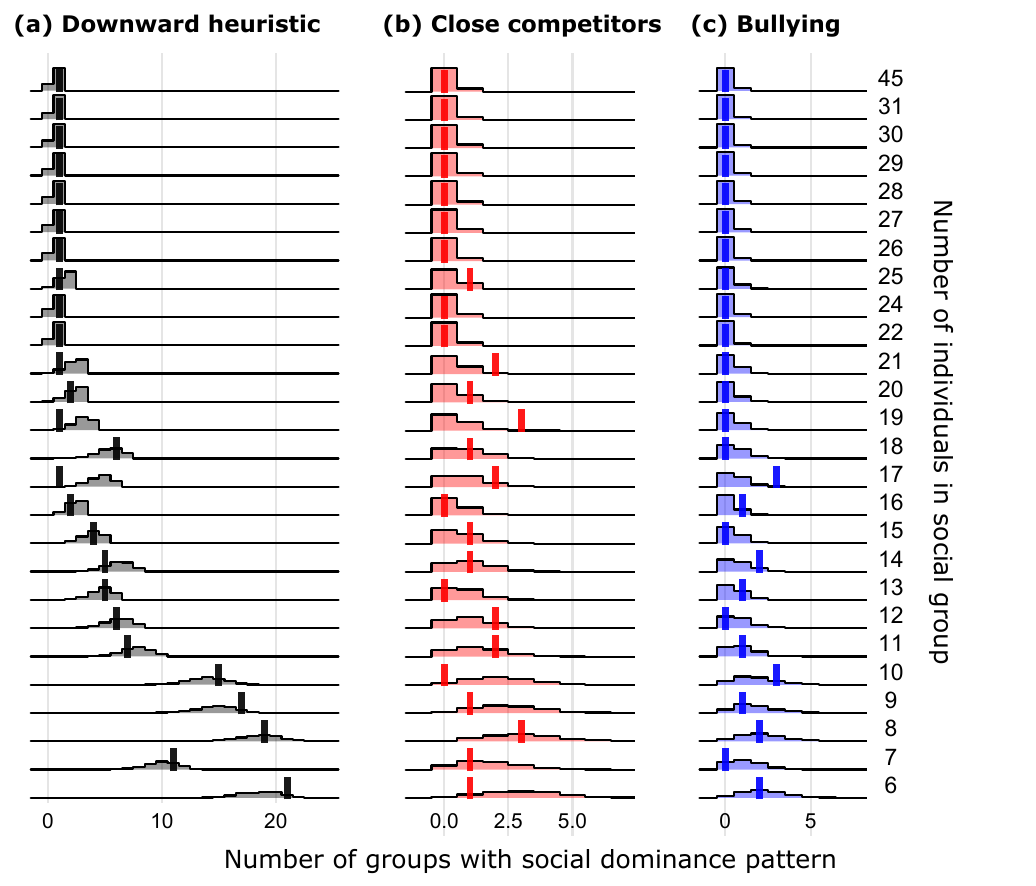}
\end{figure}

We find that in almost all cases, the observed occurrences of social dominance occurrences fall within the distributions of the randomized strategies, demonstrating that neither observation conditions (Fig.~\ref{ridgeplot_grptype}) nor group size (Fig.~\ref{ridgeplot_grpsize}) can be used to explain the observed social dominance patterns. It is important to note that these observed datasets are likely biased towards smaller group sizes and conditions where data collection is more manageable.

\clearpage
\section{The emergence of social dominance patterns with winner and loser effects} \label{SI-WLeffects}
\setcounter{figure}{0}
\renewcommand{\thefigure}{S\arabic{section}.\arabic{figure}} 
\setcounter{table}{0}
\renewcommand{\thetable}{S\arabic{section}.\arabic{table}} 

In this section, we investigated whether social dominance patterns can emerge from systems in which individuals only have information about themselves and the outcomes of their past interactions, rather than any information about their own rank or the identities or ranks of their opponents. We use this approach as a test of our logic, that more information-rich social dominance strategies only reliably emerge when individuals have some information not just about themselves, but about others in their group. We constructed several variants of a winner/loser effects model and then tested how often different social dominance patterns emerged. In the model, each individual only has access to its own win/loss record, and can only adjust its behavior based on outcomes of these events --- individuals do not have any information about which other individuals they interacted with or which individuals they have won or lost against. We draw the aggressiveness, i.e., the initial attack probability, of each individual from a uniform distribution on the interval $[0, 1]$.  

We modeled nine variants of this model (see Table~\ref{tab:effect_combos}). We investigated a winner-effect only model, a loser-effect only model, a mixed winner and loser effect model, and a model with neither a winner or loser effect. Each of these models was further investigated by incorporating each winner/loser effect as either a transient effect or persistent effect on individual behavior. We used estimates of winner and loser effect strengths from the literature~\citep{Rutte2006WhatLosing} plus a more extreme and more moderate value for comparison.

\begin{table}[htb]
\caption{The nine combinations of loser and winner effects with both transient and persistent effects. The NN combination has neither a loser nor a winner effect.}
\begin{center}
\begin{tabular}{@{}lccc@{}}
\toprule
& \multicolumn{3}{c}{\textbf{Winner effect}}\\
\cmidrule(){2-4}
\textbf{Loser effect}   & None & Transient & Persistent \\ \midrule
None       & NN   & NT        & NP         \\
Transient  & TN   & TT        & TP         \\
Persistent & PN   & PT        & PP         \\ \bottomrule
\end{tabular}
\end{center}
\label{tab:effect_combos}
\end{table}

Since individuals are not aware of their own and others' rank, the baseline ``strategy'' will be to select a random opponent and decide whether to attack based on an inherent level of aggressiveness corresponding to an initial attack probability $p_{i,0}$ for individual $i$. Computationally it is more convenient to work with odds than probabilities for the implementation of effects of past performance. So instead of the probability, we can use the initial odds of attacking, viz.
$$
    O_{i,0} = \frac{p_{i,0}}{1-p_{i,0}}
$$

To account for an individual's own past performance we implemented the following simple loser and winner effects, summarized in Table~\ref{SI-table_effects}.
    
\begin{table}[htb]
\renewcommand*{\arraystretch}{1.4}
\caption{Summary of effect types and the ways these effects are incorporated into the model behavior.}
\label{SI-table_effects} 
\begin{tabular}[t]{p{4.5cm}p{11cm}}
\textbf{Effect type }            &   \textbf{Model behavior} \\
\hline 
Transient loser effect    &   Changes the odds of attacking to $O_i = O_{i,0}\cdot \alpha_{L,t}$, where $\alpha_{L,t} < 1$, if the latest encounter resulted in a loss \\
Transient winner effect    &   Changes the odds of attacking to $O_i = O_{i,0}\cdot \alpha_{W,t}$, where $\alpha_{W,t} > 1$, if the latest encounter resulted in a win\\
Persistent loser effect &  Changes the odds of attacking by a factor $\alpha_{L,p} < 1$ every time an individual loses \\
Persistent winner effect    & Changes the odds of attacking by a factor $\alpha_{W,p} > 1$ every time an individual wins\\   
  \hline
\end{tabular}
\end{table}

Note that the transient effects are not cumulative, \emph{i.e.}, they change the odds relative to the initial odds for one encounter only. The persistent effects, on the other hand, change the odds relative to what they were in the previous encounter, \emph{e.g.}, two consecutive losses will multiply (decrease) $O_i$ by a factor $\alpha_{L,p}^2$. To avoid unrealistically low or high odds of attacking a lower limit of $\alpha_\text{min} = 10^{-3}$ and an upper limit of $\alpha_\text{max} = 10^3$ are imposed.

In the simulations we have attributed random initial aggression levels, $p_{i,0}$, but fixed the $\alpha$ factors controlling the winner and loser effects to the same value across all individuals.

We used three different sets of $\alpha$ values in our simulations: (a) \textbf{extreme values} ($\alpha_L = 0.05$ and $\alpha_W = 2.7$), (b) \textbf{realistic values} ($\alpha_L = 0.18$, $\alpha_W = 1.87$, which are the pooled estimates from the meta-analysis by Rutte et al.~\cite{Rutte2006WhatLosing}), and (c) \textbf{moderate values} ($\alpha_L = 0.7$ and $\alpha_W = 1.3$). As a control we ran simulations with the same parameters but without any of the winner or loser effects. In all cases, we used a group size of $N=10$, and let the simulation run for a total of 1000 attacks.

\begin{table}[htb]
\caption{Frequency of the observed social dominance patterns for the each of the three parameter sets and for both transient and persistent winner and loser effects: (a) extreme values, $\alpha_L = 0.05$ and $\alpha_W = 2.7$, (b) realistic values, $\alpha_L = 0.18$, $\alpha_W = 1.87$ (which are the pooled estimates from the meta-analysis by Rutte et al.~\cite{Rutte2006WhatLosing}), (c) moderate values, $\alpha_L = 0.7$ and $\alpha_W = 1.3$, and (d) the sum total of runs across all models a-c that exhibited each social dominance pattern.}
    \centering
    \resizebox{\textwidth}{!}{
\begin{tabular}{lcccccccc}
 & \multicolumn{2}{c}{(a)} & \multicolumn{2}{c}{(b)}  & \multicolumn{2}{c}{(c)}  & \multicolumn{2}{c}{(d)} \\
 & \multicolumn{2}{c}{\textbf{Extreme $\alpha$ values}} & \multicolumn{2}{c}{\textbf{Rutte et al. $\alpha$ values}} & \multicolumn{2}{c}{\textbf{Moderate  $\alpha$ values}} & \multicolumn{2}{c}{\textbf{Summary (all $\alpha$ values)}} \\ \cmidrule(lr){2-3}\cmidrule(lr){4-5}\cmidrule(lr){6-7}\cmidrule(lr){8-9}
 & Transient & Persistent & Transient & Persistent & Transient & Persistent & Transient & \multicolumn{1}{c}{Persistent} \\ 
 
\textbf{Social dominance pattern}  & \textit{n} & \textit{n} & \textit{n} & \textit{n} & \textit{n} & \textit{n} & \textit{n} & \multicolumn{1}{c}{\textit{n}} \\ 
\midrule
Bully  & $\phantom{00}0$ & $\phantom{00}0$ & $\phantom{00}0$ & $\phantom{00}2$ & $\phantom{00}0$ & $\phantom{00}0$ & $\phantom{00}0$ & $\phantom{00}2$ \\
Close competitors  & $\phantom{00}5$ & $\phantom{00}4$ & $\phantom{00}2$ & $\phantom{00}3$ & $\phantom{00}7$ & $\phantom{00}7$ & $\phantom{0}14$ & $\phantom{0}14$ \\
Downward heuristic  & $\phantom{0}87$ & $\phantom{0}90$ & $\phantom{0}83$ & $\phantom{0}90$ & $\phantom{0}80$ & $\phantom{0}83$ & $250$ & $263$ \\
Undefined  & $\phantom{00}8$ & $\phantom{00}6$ & $\phantom{0}15$ & $\phantom{00}5$ & $\phantom{0}13$ & $\phantom{0}10$ & $\phantom{0}36$ & $\phantom{0}21$ \\
All  & $100$ & $100$ & $100$ & $100$ & $100$ & $100$ & $300$ & $300$ \\
\bottomrule 
\end{tabular}

    }
    \label{tab:strat_freq}
\end{table}

\begin{figure}[htb]
    \centering
        \caption{The distribution of focus and position for each of the 100 groups simulated for each combination of parameters ($\alpha_W, \alpha_L$) and each of the two types of winner and loser effects (transient and persistent). The $\alpha_W, \alpha_L$ combinations used are consistent with Table~\ref{tab:strat_freq}: extreme values where $\alpha_L = 0.05$ and $\alpha_W = 2.7$, Rutte et al. indicates ``realistic'' values where $\alpha_L = 0.18$, $\alpha_W = 1.87$ (which are the pooled estimates from the meta-analysis by Rutte et al.~\cite{Rutte2006WhatLosing}), and moderate values where $\alpha_L = 0.7$ and $\alpha_W = 1.3$). The group-level social dominance pattern is indicated by the color of the point representing each group.}
    \label{fig:FP_facet}
    \includegraphics[width=0.9\textwidth]{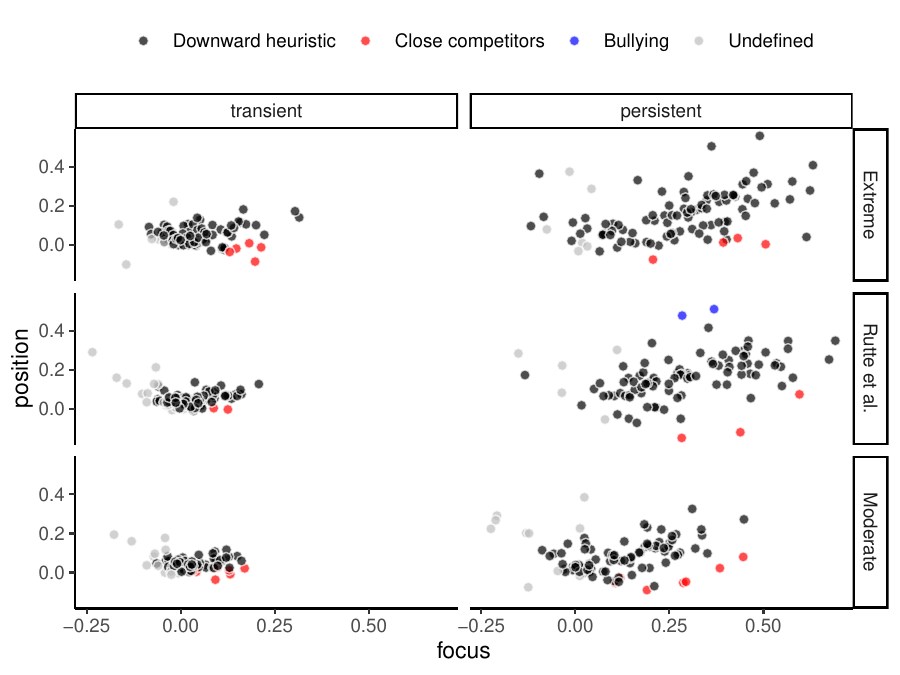}

\end{figure}

\begin{figure}[htb]
    \centering
    \caption{Focus and position values for the simulated groups using the ``realistic'' values of $\alpha_L = 0.18$, $\alpha_W = 1.87$ (the pooled estimates from the meta-analysis by Rutte et al.~\cite{Rutte2006WhatLosing}). \emph{Left:} All 100 groups. \emph{Right:} Same data, but subset to show only groups with focus $\geq 0$ and position $\geq 0$.}
    \label{fig:Rutte_FP}
    \includegraphics[width=0.49\textwidth]{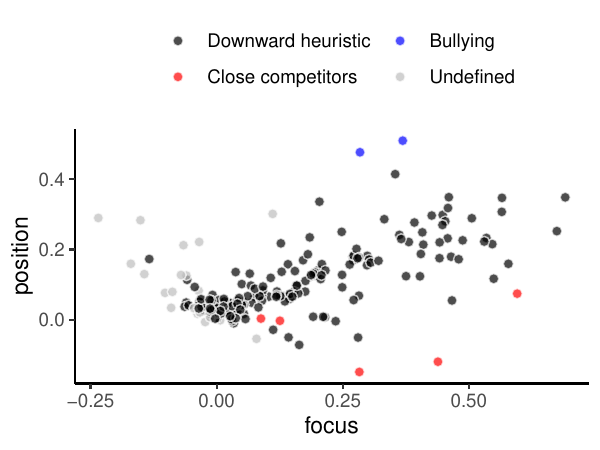}
    \includegraphics[width=0.49\textwidth]{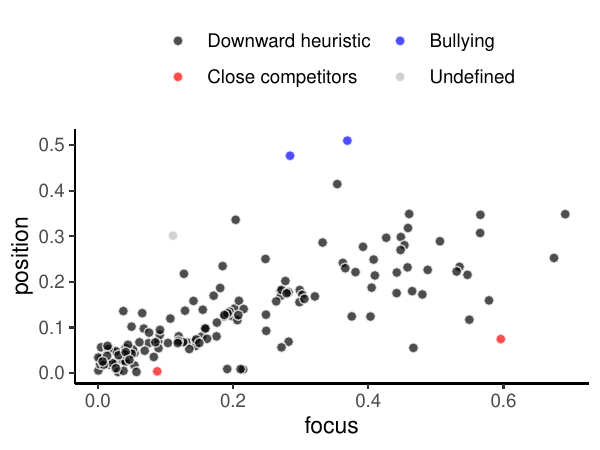}
\end{figure}

Some parts of the structure exhibited by our generative models, using only winner and/or loser effects, differed from the structure of the observed social groups in our hierarchy dataset (Section~\ref{SI-strucagg}). For example, many more of the generated groups show focus and/or position values less than zero (Fig.~\ref{fig:FP_facet}), which was not often observed in the empirical groups. To highlight this difference, we examined one of the parameter sets (the ``realistic'' values of $\alpha_L = 0.18$, $\alpha_W = 1.87$ from Rutte et al., for persistent effects). The plot of the full data set with all 100 groups is shown on the left in Figure~\ref{fig:Rutte_FP} and the censored data including only non-negative values is shown on the right.

Using this censored dataset (which excludes artificial groups which which showed characteristics inconsistent with our empirical datasets), we found that the two more complex social dominance patterns were rarely produced by winner and/or loser effects. Transient effects resulted in 0\% bully and $<2\%$ close competitor patterns and persistent effects resulted in $<3\%$ bully and just over 1\% close competitor patterns (Table~\ref{tab:Rutte}). For uncensored data, bully and close competitor patterns were still rare: across all combinations of $\alpha_L$ and $\alpha_W$ (including groups modeled with moderate, realistic, and extreme values) transient effects produced 0 groups with a bullying pattern and $4.7\%$ of groups (14 groups total) with a close competitor pattern (Table~\ref{tab:strat_freq}d). Even considering the most extreme $\alpha_L$ and $\alpha_W$ alone (Table~\ref{tab:strat_freq}a), our models produced no groups with a bullying pattern and just 5\% (transient effects) and 4\%(persistent effects) with close competitors patterns. 

These results show that bullying or close competitors social dominance patterns are rarely produced through winner and/or loser effects alone. These results provide additional justification for our argument that additional social information, beyond an individual merely remembering the outcomes of its own encounters, is generally needed to produce the more complex social dominance patterns.

\begin{table}[htb]
    \centering
    \caption{Prevalence of the different social dominance patterns in the censored data set for the Rutte et al. values of $\alpha_W$ and $\alpha_L$ (values summarized from Fig.~\ref{fig:Rutte_FP}, right (persistent effects), with additional summaries for transient effects). The equivalent numbers for the uncensored data are shown in Table~\ref{tab:strat_freq}b.}
    \label{tab:Rutte}
    \begin{tabular}{lcccc}
 & \multicolumn{2}{c}{Number of groups} & \multicolumn{2}{c}{Percent of groups} \\ \cmidrule(lr){2-3}\cmidrule(lr){4-5}
Social dominance pattern  & Transient & Persistent & Transient & \multicolumn{1}{c}{Persistent} \\ 
\midrule
Bully  & $\phantom{0}0$ & $\phantom{0}2$ & $\phantom{00}0.00$ & $\phantom{00}2.27$ \\
Close competitors  & $\phantom{0}1$ & $\phantom{0}1$ & $\phantom{00}1.82$ & $\phantom{00}1.14$ \\
Downward heuristic  & $54$ & $84$ & $\phantom{0}98.18$ & $\phantom{0}95.45$ \\
Undefined  & $\phantom{0}0$ & $\phantom{0}1$ & $\phantom{00}0.00$ & $\phantom{00}1.14$ \\
All  & $55$ & $88$ & $100.00$ & $100.00$ \\
\bottomrule 
\end{tabular}

\end{table}


\end{document}